\documentstyle[12pt,aasms4]{article}
\def\mass{{\hbox{\cal M}}}
\def\MSun{{\hbox{\mass$_\odot$}}}
\def\ltsim{\, {}^<_\sim \,}
\def\etal{{\it et al.}}
\def\ie{{\it i.e.}}
\def\eg{{\it e.g.}}

\def\dv{$\Delta V_{\rm TO}^{\rm HB}$}

\def\deg{\ifmmode^\circ\else$^\circ$\fi}    

\def\hper{\ifmmode \rlap.^{h}\else $\rlap{.}^h$\fi} 
\def\sper{\ifmmode \rlap.^{s}\else $\rlap{.}^s$\fi}    

\def\deg{${}^\circ$}

\def\sec{${}^{\prime\prime}$}
\def\gtsim{ \,{}^>_\sim\, }
\def\hst{{\it HST\/}}
\def\vmi{\hbox{\it V--I\/}}
\def\bmv{\hbox{\it B--V\/}}
\def\m#1{M$\,$#1}
\def\mmm{\hbox{\rm (m--M)}}
\def\ngc#1{NGC$\,$#1}
\def\margin#1{\marginpar{\hfill \scriptsize #1\ }}
\def\today{\number\year\space \ifcase\month\or
  January\or February\or March\or April\or May\or June\or
  July\or August\or September\or October\or November\or December\fi
  \space\number\day}
\def\now{\number\year\space \ifcase\month\or
  January\or February\or March\or April\or May\or June\or
  July\or August\or September\or October\or November\or December\fi
  \space\number\day .\number\time}
\reversemarginpar

\lefthead{Stetson et al.}
\righthead{Outer Halo Clusters}

\begin{document}

\title{Ages for Globular Clusters in the Outer Galactic Halo: The Second
Parameter Clusters Palomar~3, Palomar~4 and Eridanus\footnote{ Based on
observations made with the NASA/ESA {\it Hubble Space Telescope}, obtained
at the Space Telescope Science Institute, which is operated by the
Association of Universities for Research in Astronomy, Inc., under NASA
contract NAS5-26555.}}

\author{Peter B. Stetson}
\affil{Dominion Astrophysical Observatory, Herzberg Institute of
Astrophysics, National Research Council, 5071 West Saanich Road,
Victoria, British Columbia V8X 4M6, Canada
\\Electronic mail: peter.stetson@hia.nrc.ca}
 
\author{Michael Bolte}
\affil{UCO/Lick Observatory and the Department of Astronomy
and Astrophysics, University of California, Santa Cruz,
California 95064
\\Electronic mail: bolte@lick.ucolick.org}
 
\author{William E. Harris}
\affil{Department of Physics, McMaster University, Hamilton, Ontario L8S 4M1,
Canada
\\Electronic mail: harris@physics.mcmaster.ca}
 
\author{James E. Hesser and Sidney van den Bergh}
\affil{Dominion Astrophysical Observatory, Herzberg Institute of
Astrophysics, National Research Council, 5071 West Saanich Road,
Victoria, British Columbia V8X 4M6, Canada
\\Electronic mail: james.hesser@hia.nrc.ca, sidney.vandenbergh@hia.nrc.ca}

\author{Don~A.~VandenBerg}
\affil{University of Victoria, Department of Physics \& Astronomy,
       PO Box 3055, Victoria, British Columbia  V8W 3P7
\\Electronic mail: davb@uvvm.uvic.ca}

\clearpage

\author{Roger~A.~Bell}
\affil{University of Maryland, Department of Astronomy,
       College Park MD 20742-2421
\\Electronic mail: rabell@astro.umd.edu}

\author{Jennifer~A.~Johnson}
\affil{UCO/Lick Observatory and the Department of Astronomy
and Astrophysics, University of California, Santa Cruz,
California 95064
\\Electronic mail: jennifer@lick.ucolick.org}
 
\author{Howard~E.~Bond and Laura~K.~Fullton}
\affil{Space Telescope Science Institute, 3700 San Martin Drive,
       Baltimore MD 21218 
\\Electronic mail: bond@stsci.edu, fullton@stsci.edu}
 
\author{Gregory G.~Fahlman and Harvey B.~Richer}
\affil{Department of Physics \& Astronomy, University of British
       Columbia, Vancouver, British Columbia V6T 1Z4 
\\Electronic mail: richer@astro.ubc.ca, fahlman@astro.ubc.ca}
 
\vspace{2 in}
\begin{abstract}

We have used the WFPC2 camera on the {\it Hubble Space Telescope\/}  to
obtain photometry of the outer-halo globular clusters Palomar~3,
Palomar~4, and Eridanus. These three are classic examples of the ``second
parameter'' anomaly because of their red horizontal-branch morphologies in
combination with their low-to-intermediate metallicities.  Our
color-magnitude diagrams in $(V, \vmi)$ reach $V_{lim} \simeq 27.0$,
clearly delineating the subgiant and turnoff regions and about three
magnitudes of the unevolved main sequences.  The slopes and dereddened
colors of the giant branches are consistent with published [Fe/H]
estimates that rank the clusters (Pal~3, Eridanus, Pal~4) in order
of increasing metallicity, with all three falling near or between the
abundance values of the classic nearby halo clusters \m3 and \m5.
Differential fits of their color-magnitude diagrams are made to each other
and to \m3 and \m5 for relative age determinations.  We find that the
three outer-halo cluster CMDs differ from the nearby clusters in a way
that is consistent with their being younger by $\sim 1.5 - 2\,$Gyr, if we
have correctly estimated the clusters' chemical-abundance ratios.
Conversely, the inferred age difference could be smaller ($\ltsim 1\,$Gyr)
if either [Fe/H] or [$\alpha$/Fe] for the outer-halo clusters is
significantly lower than we have assumed.  Possible age spreads of order
1$\,$Gyr among both the nearby and outer-halo clusters may also be
present.

\end{abstract}


\section{Introduction}

This is the second paper in which we report results of age measurements
for the globular clusters at the outermost reaches of the Milky Way
halo.  In the first paper of this series (Harris \etal\ 1997 [Paper I]),
we showed that the massive globular cluster \ngc{2419}, at a
Galactocentric distance of $\sim 90$ kpc, has a color-magnitude diagram
(``CMD'') that is indistinguishable from that of \m92 and other nearby
halo clusters with the lowest known metallicities, [Fe/H] $\simeq -2.2$.
These clusters have the same age to within the measurement precision of
$\simeq 1$ Gyr, suggesting that globular cluster formation may have {\it
started\/} at virtually the same time everywhere across the entire expanse
of the protoGalaxy.

However, \ngc{2419} represents only the upper envelope of the chronology
of the outermost-halo clusters.  The other remote clusters (Palomar~3, 4,
14, Eridanus, and AM-1) are intriguing in quite a different way as strong
cases of the classical ``second parameter'' problem:  their
horizontal-branch (``HB'') morphology is redder than the mean for nearby
halo clusters of similar (low to intermediate) [Fe/H].  The tendency for
the second-parameter effect to become increasingly evident in the outer
halo is one of the key motivations behind the Searle \& Zinn (1978 [SZ])
scenario in which the Galaxy's spherical halo is envisaged as having
accumulated over time in a series of mergers of sub-galactic fragments.
Making the plausible assumption that age is the most important second
parameter responsible for intrinsic scatter in the relation between [Fe/H]
and HB morphology, SZ concluded that the mean age of the clusters
decreases, and the age spread increases with increasing Galactocentric
distance, $R_{\rm GC}$.  By contrast, if all parts of the spherical halo
formed into stars during a rapid free-fall collapse as was argued by
Eggen, Lynden-Bell \& Sandage (1962 [ELS]), it would be much more likely
that the halo clusters would have rather similar ages and would (for
example) follow a simple relation between CMD morphology and metallicity.

Arguing from a comparison of quantitative simulations of HB morphology as
a function of metal abundance and age to CMDs of real clusters Lee,
Demarque \& Zinn (1994, hereinafter LDZ; see especially their Fig.~7)
concluded that the globular clusters within $\sim\,8\,$kpc of the Galactic
center --- spanning the full metal-abundance range from less than one
one-hundredth solar to nearly the solar value --- are essentially coeval.
Those at Galactocentric distances between 8 and 40$\,$kpc formed an
average of some 2$\,$Gyr later, but with a perceptible dispersion in their
formation epochs.  Finally, the five Galactic globulars with red
horizontal branches that lie outside 40$\,$kpc formed more than 4$\,$Gyr
later than their counterparts in the Galactic center (these statements are
based upon our interpretation of LDZ's Fig.~7).

Lee (1992) has added another wrinkle to the earliest formation history of
the Galaxy, using another argument based on HB morphology.  He notes that
the RR~Lyrae variables in Baade's window --- a region where we are allowed
a more or less clear glimpse into the Galactic bulge --- possess metal
abundances closely bunched about a value of [Fe/H] = --1.0.  He then
argued that for stars of such high metal abundance to populate the HB
instability strip, they must be $\sim\,1\,$Gyr older than even the mean
age of the globular clusters with $R_{\rm GC} < 8\,$kpc.

However, the foregoing arguments rely on the assumptions that age is the
{\it dominant\/} second parameter after metal abundance, and that HB
morphology can be used in a simple and direct way as a measure of cluster
age.  These assumptions are straightforward and widely accepted, but they
are perhaps not yet ironclad, and the debate continues (\eg, Stetson,
VandenBerg \& Bolte 1996; Sarajedini, Chaboyer \& Demarque 1997).  There
are two additional steps required to test and calibrate this chronology
directly:  first, we must measure in considerable detail the chemical
compositions of stars in these clusters (since composition differences can
mimic age differences in a variety of ways); and second, we must make much
more direct estimates of age based on photometry of the main sequence and
subgiant branch in clusters spanning the range of Galactic environments.
Stars in these stages of evolution are driven by somewhat simpler internal
physics, and suffer from fewer potential complications due to uncertain
mid-life processes such as mixing and mass-loss, than the HB stars (see,
\eg, Sweigart 1997).

Among the six Galactic globular clusters at Galactocentric distances
greater than (to choose an arbitrary boundary) that of the Large
Magellanic Cloud, only one --- Pal~14 --- has had its main-sequence
turnoff detected from the ground (Sarajedini 1997), although even in this
best case the data penetrate less than a magnitude past the turnoff.
Sarajedini argues that Pal~14, with [Fe/H] $\approx$ --1.6, is 2 Gyr
younger than the nearby globular \m5 and 4 Gyr younger than NGC 6752.
Additional caveats on these results are that M5 may have a significantly
higher [Fe/H] than Pal~14, and NGC~6752 has an extremely blue HB
morphology that makes it very difficult to register it and Pal~14 to a
common distance modulus.

It is possible that the
outermost-halo objects may actually bear little direct relevance to the
formation history of the main part of the Galaxy.  Many or all of the
red-HB clusters in particular (Pal~3, 4, 14, Eridanus, AM-1) may belong to
a stream of remote satellites (the Fornax-Leo-Sculptor stream; see
Majewski 1994) in an orbital plane with an extremely large semimajor axis,
$R \sim 150$ kpc or more.  No matter whether the Galaxy itself formed in a
monolithic {\it or\/} a hierarchical fashion, it is possible that this
handful of objects was born  well apart from the original protoGalaxy,
and are now left orbiting a large stellar system of which they were never
really an integral part.  If this is true, these clusters may merely be a
red herring in the search for clues to the formation of the Milky Way.
But at least these same clusters should still provide us with a test of
the proposed correlation between HB morphology and age, which is currently
regarded as the essence of the second-parameter problem.  In this paper,
we use data from the {\it Hubble Space Telescope\/} to address the
questions of {\it whether\/}, and {\it to what extent\/}, the
intermediate-metallicity clusters of the outer halo and the nearby halo
really do have different ages.  (Following LDZ, we separate the ``inner''
halo from the ``intermediate'' halo at a Galactocentric distance of about
$R_\odot$, and the ``intermediate'' from the ``outer'' halo at about
50$\,$kpc.  In this paper, we use the word ``nearby'' in its common
English sense, to describe those clusters that are close to the Sun and,
hence, usually quite well studied.  It encompasses some of the outermost
members of the inner halo as well as the closer members of the
intermediate halo; ``nearby clusters'' without the additional modifier
``halo'' can also include a few of the nearer disk and/or bulge clusters.)

\section{Observations and Data Reduction}

The clusters we report on here are Palomar~3, Palomar~4,\footnote{Results
from reliminary analyses of the present data for Pal$\,$3 and Pal$\,$4 were
included in the study by Richer \etal\ (1996).  The current analysis
supersedes the results cited in that paper.} and Eridanus.  These are
three of the six known globulars beyond a Galactocentric distance $R_{\rm
GC} \sim 50$ kpc, and (typically for such objects) have low luminosities,
diffuse structures, and low to intermediate [Fe/H] values.  Pal~3 and
Pal~4 were imaged with the WFPC2 camera during \hst\ Cycle 4, with Pal~4
being visited twice.  Eridanus was observed during Cycle 5.  As was the
case for \ngc{2419}, the first visit to Pal~4 resulted in eight long
exposures in F555W (each of 1400$\,$s duration) and F814W (of 1300$\,$s),
after which the spacecraft was shifted by 20\sec\ toward the center of the
cluster, and six short exposures were taken in each filter (60$\,$s in
F555W and 30$\,$s in F814W).  For the second visit to Pal~4 and the visit
to Pal~3, a different scheme was employed.  For these observations we
specified that the cluster be centered on the PC1 chip and requested
cycles of exposures, each cycle consisting of an (approximately)
1800$\,$s, a 60$\,$s, and a 30$\,$s exposure; we obtained eight such
cycles per filter for each cluster.  Because Eridanus is closer than the
other two clusters, for this system we obtained seven exposures of 1100 or
1200$\,$s duration in each of F555W and F814W, plus $3\times180\,$s in
F555W and $3\times230\,$s in F814W.  Extensive additional details about
these observations may be obtained by electronic query to either the Space
Telescope Science Institute in Baltimore (http://archive.stsci.edu/HDA/),
or the Canadian Astronomy Data Centre
(http://cadcwww.hia.nrc.ca/hst/science.html; specify GO programs 5481,
5672, and 6106: J.~E.~Hesser, PI), or by contacting the first author of
this paper.

The process of data reduction was the same as that for \ngc{2419} in
Paper~I, and will be described more completely in Stetson (1998, in
preparation).  The suite of codes collectively referred to as the
third-generation DAOPHOT (including ALLFRAME; see Stetson 1994) was
employed to obtain instrumental magnitudes and colors for all measurable
stars in the WFPC2 fields and to merge these onto a common photometric
system.  Raw observed instrumental magnitudes were corrected for charge
loss due to charge-transfer inefficiency by an amount equal to
--0.04$\,$mag per 800$\,$pixels times the $y$-coordinate of the star on
the chip (Holtzman \etal\ 1995; Whitmore \& Heyer 1997).  Color terms for
the transformation from the instrumental (F555W, F814W) magnitude system
to the standard Johnson $V$, Kron-Cousins $I$ systems were taken directly
from Holtzman \etal\ (their Table~7).  The zero-point calibration relies
ultimately on ground-based images of some of these same fields obtained by
us.  We found that the zero points inferred for our {\it short\/}
exposures agree with those from Holtzman \etal\ to within $\pm 0.02$ mag
in both $V$ and $I$.  However, the zero points appropriate to our long
exposures appear to differ from those of Holtzman \etal\ by some 0.05~mag
in each filter, in the sense that the effective quantum efficiency of the
CCDs in WFPC2 appears to be higher in long exposures than in short ones.
We believe that this behavior is consistent with the notion that
crystal-lattice imperfections in the CCDs act as charge traps.  In long
exposures, where the diffuse sky foreground is appreciable, these traps
consume electrons and lower the overall diffuse flux in the image without
affecting stellar profiles or brightnesses.  However, in short exposures
the diffuse sky is not sufficient to fill all the traps, and electrons are
consumed from the charge packets representing stellar images either during
the exposure itself or during readout of the chip.  This effect appears to
operate {\it in addition\/} to the charge-transfer inefficiency ramp of
--0.04$\,$mag per 800 pixels mentioned above.  Fortunately, for the
differential comparisons to be performed below accurate knowledge of the
zero points is not essential:  incorrect values will slightly falsify the
distances and/or reddening values inferred for our clusters, but the
relative age indicators will be unaffected.  We hope to be able to improve
the fundamental accuracy of our zero points once the charge-transfer
inefficiency of the WFPC2 CCDs becomes better characterized.  For our
present purposes, separate photometric zero-points for the long and short
exposure times have been determined by direct comparison to our
ground-based observations of Pal~4 and \ngc{2419}.

Later, we will compare the results for these three clusters to our
ground-based photometry for the nearby halo clusters \m3 and \m5.  Data
for \m3 were obtained during seven observing runs on four different
telescopes between 1983 and 1994 (although the $I$-band data came from
only two of those runs), while \m5 was observed during twelve runs on six
different telescopes during the same time span (again the $I$-band data
were from two of those runs).  These data include the observing run
analysed by Johnson \& Bolte (1998), but the images have been
independently reanalysed and recalibrated for the present study.  All the
data from the various runs were homogenized to a common photometric system
--- that of Landolt (1992) --- as originally outlined some years ago by
Stetson (1990, 1993).  The present ground-based photometric system is
anchored to a total of some 44 observing runs in $V$ and 17 runs in $I$.

\section{The Color-Magnitude Diagrams}

Figures 1 through 3 \margin{Figs.~1--3} show the CMDs we have derived from
the \hst\ imagery for the remote clusters, Pal~3, Pal~4, and Eridanus.
For each object, the data from the four WFPC2 chips have been combined.
The narrowness of the red giant branches (``RGBs'') in all three CMDs is
consistent with our expectation that the four WFPC2 CCDs and the long and
short exposures have all been normalized to the same photometric system to
within $\sim\pm0.01\,$mag.  We note that all three of the outer-halo clusters
contain significant numbers of blue straggler stars (about a dozen such
objects appear in each of Figs.~1--3).  Figures~4 and 5\margin{Figs. 4 \&
5} illustrate the ground-based color-magnitude diagrams for \m3 and \m5
that we will later compare to those of the outer-halo clusters.  The
present data for these clusters give results that are very similar to
those of Johnson \& Bolte (1998), with about twice the sample size.

Palomar~3 is the only outer-halo cluster in our study which appears to
contain RR Lyrae variables.  Three were found by Gratton \& Ortolani
(1984), who also identified one candidate Population~II Cepheid; a fifth
variable candidate, which had previously been noted by Burbidge \& Sandage
(1958), was considered to be non-variable by Gratton and Ortolani.  One of
Gratton \& Ortolani's RR~Lyrae stars, their number 34, lies outside our
field coverage, as does the candidate Cepheid (their number 102).  We have
recovered the other two candidate RR Lyraes from Gratton and Ortolani;
their number 155 = our number 1-233, and their number 283 = our number
2-614.  In addition, in contrast to the conclusion of Gratton and
Ortolani, we find that the Burbidge-Sandage candidate = Gratton-Ortolani
number 188 = our number 2-198, is indeed a real variable.  Finally, we
identify four additional RR~Lyrae candidates not previously noted.  In
Figure~6\margin{Fig.~6} we present light-curve fragments from our data for
these seven stars.  Within the limitations of the available data, all
seven of the variables appear to have the asymmetric sawtooth lightcurve
characteristic of RR$_{ab}$--type variables.  Table~1\margin{Table~1}
represents notional period estimates for the stars from these data (for
our present crude purposes, we neglect the slight difference in phase
between extrema as observed in the two photometric bandpasses).  We think
it most likely that for five of the seven candidate variables, three full
cycles occurred between the two dates of observation, while for the other
two, candidates 1-299 and 2-614, barely more than two cycles elapsed
between the end of the first observing sequence and the start of the
second.  These assumptions yield virtually identical periods of 0.6$\,$d
for all seven of the RR~Lyrae candidates.  The possibility of four cycles
having occurred for any of the variables, implying a period of order
0.44$\,$d, can be ruled out by the absence of significant lightcurve
overlap on either occasion, since on each date we obtained continuous data
strings more than 0.48$\,$d in length.  Conversely, the possibility that
some of the variable candidates underwent only two full cycles or less
during the course of observations would imply periods of order 0.9$\,$d,
which are extremely rare among classical RR~Lyraes.  The mean magnitude
and color of all seven candidates is indicated in Fig.~1 as $\langle V
\rangle = 20.49 \pm 0.027$, $\langle \vmi \rangle = 0.66 \pm 0.061$.
Among the non-variable stars in our Pal~3 sample, we count six in the
neighborhood of the red horizontal branch, implying a horizontal-branch
morphology index for this cluster of $(B-R)/(B+V+R) = -6/13 = -0.5\pm0.2.$

We have determined fiducial points for the principal CMD sequences in two
ways.  First, we simply calculated robust mean magnitudes and colors
(after three iterations of sigma-clipping) in bin steps of $\Delta V
\simeq 0.2\,$mag for the three outer-halo clusters, and in steps of
0.1$\,$mag for M3 and M5.  These mean points are shown as squares in
Figures 7 through 11 and are listed in Table 2:  columns 1 and 2 give the
mean $(V,V-I)$ values for the stars found in each bin, while the last
three columns give the number of stars per bin retained in each of the
three iterations.  Notice that the bin steps are not strictly uniform in
$\Delta V$, particularly along the upper giant branch, where the bin
intervals and sizes are determined more strongly by the (patchy)
distribution of RGB stars.

Second, we took D.~A.~VandenBerg's latest set of theoretical isochrones
and, regarding them as a set of numerical drafting splines likely to
resemble an actual cluster locus more closely than, say, a low-order
polynomial, fitted them directly to the original data.  These isochrones
are essentially the ones described in Paper~I except that they now
incorporate the latest neutrino cooling rates (Itoh \etal\ 1996), which
result in a small increase in the derived core mass at the giant-branch
tip (0.002--0.004$\,\MSun$) and an increase of $\sim\,$0.025$\,$mag in the
corresponding HB luminosity.  These changes are irrelevant for our
immediate purposes.  Some complications, such as the uncertain effects of
convection and mixing on the derived radii for evolved stars, reddening,
and the practical difficulty of accurately characterizing the relationship
between colors on the Johnson/Kron-Cousins standard system and those
obtained with a particular filter/detector combination, continue to
constitute problems for the reliable {\it absolute\/} interpretation of
CMDs (\eg, VandenBerg 1998, in preparation).  The magnitude and color
zero-points of the isochrones employed here are based on the use of
theoretical zero-age HB models as absolute luminosity fiducial points, but
it turns out that our models are numerically consistent with a compromise
between cluster distance estimates from the latest Hipparcos parallaxes
for subdwarfs on the one hand (\eg, Pont \etal\ 1998), and for RR~Lyraes
on the other (Fernley \etal\ 1998).  Clearly, absolute cluster ages will
be quite a strong function of one's assumptions concerning the distance
scale (see, \eg, Chaboyer \etal\ 1998 and Gratton \etal\ 1997; also,
Hendry 1997), but unless we are very unlucky age {\it differences\/} from
direct cluster-to-cluster comparisons will be minimally affected.

We used theoretical isochrones for [Fe/H] = --1.14, --1.31, --1.41,
--1.54, and --1.61; [$\alpha$/Fe] = 0.0, +0.3, and +0.6; and age = 8, 10,
12, 14, 16, and 18~Gyr.  We numerically fitted {\it all\/} of these
isochrones to the data for {\it each\/} of the clusters, allowing
arbitrary vertical and horizontal shifts in each case.  Isochrones were
fitted to the actual cluster data (\ie, {\it not\/} to the normal points),
and the fits were carried out in the $(I,\vmi)$ plane, because in this
representation the subgiant branch is more strongly sloped than in the
$(V,\vmi)$ diagram.  Looked at in another way, if apparent magnitude is
taken as the independent variable and color as the dependent one, then the
long, flat subgiant branch characteristic of young ages becomes vertical
in the function \vmi~=~Fn($V$); indeed, for very young ages the
relationship can be multi-valued so the mapping between color and
magnitude becomes equivocal.  The relation \vmi~=~Fn($I$) is better
behaved in this regard.  

The fits were carried out using a robust maximum-likelihood scheme that
considered the photometric uncertainty of each stellar measurement and
reduced the relative weights of obvious outliers.  This method
(essentially the same as the one illustrated in Fig.~6 of Stetson 1987) is
quite insensitive to the actual numbers and distribution of outliers, and
is also robust against variations in the details of the weighting scheme.
After all the isochrone fits had been done, the one isochrone which
produced the highest maximum of this likelihood function was identified
and adopted as representing the position and shape of the turnoff and
subgiant region of the CMD.  Note that in this procedure we imposed no
{\it a~priori\/} constraints on cluster distance, reddening, chemical
abundance, or age within the parameter space spanned by these particular
isochrones, nor do we intend to draw any immediate {\it a~posteriori\/}
conclusions concerning those parameters.  At this point, all we want is to
utilize the theoretical isochrones as a set of numerical curves describing
the shape and position of the turnoff and subgiant branch in the CMD.
That said, we note that the specific isochrones selected by the
statistical procedure were, in fact, reasonable:  the best matches were
achieved with model metal abundances in the range --1.41$\, < \,$ [Fe/H]
$\, < \,$ --1.61 and model ages $\sim\,$14--16$\,$Gyr.  We stress again
that no great weight is to be placed on these numerical quantitities,
apart from a general sense of satisfaction that they are not
unreasonable.

Having determined which isochrone best matches the appearance of the
cluster sequence within two magnitudes of the turnoff,  $I_{TO}-2 < I <
I_{TO}+2\,$mag (solid curves in Figs.~7--11), we then read out from the
isochrones the magnitudes and colors corresponding to five points:
(1)~the turnoff, (2)~the point on the subgiant branch that is precisely
0.10~mag redder than the turnoff, (3)~the point on the subgiant branch
that is 0.05~mag redder than the turnoff, and (4) and (5) the points on
the upper main sequence that are precisely 0.05 and 0.10~mag redder than
the turnoff.  After the $(I,\vmi)$ locations of these fiducial points had
been determined, the magnitudes were converted to their $V$-band
equivalents.  We present the apparent visual magnitudes of isochrone
points (2) through (5), denoted by  $V^-_{+0.10}$, $V^-_{+0.05}$,
$V^+_{+0.05}$, and $V^+_{+0.10}$, respectively, in
Table~3\margin{Table~3}.  For purposes of the present discussion, we took
the turnoff of an isochrone to be given by the magnitude $M_{I,TO}$ such
that the $(\vmi)_\circ$ color of the isochrone at a point 0.2~mag more
luminous than $M_{I,TO}$ is exactly equal to the color at a point 0.2~mag
less luminous than $M_{I,TO}$.  The tabulated color on the fitted
isochrone at $M_{I,TO}$ was then taken as the turnoff color of the
cluster.  This definition was adopted to avoid uncertainty due to minor
numerical glitches either in the theoretical evolution calculations or in
the interpolation from evolutionary tracks to isochrones.  However, we
subsequently found that the turnoff color defined in this way agreed to
within $10^{-4}$~mag of the bluest color actually achieved on the
isochrone, so it seems our concerns were exaggerated.

VandenBerg, Bolte \& Stetson (1990, ``VBS'') suggested using the
main-sequence point $V^+_{+0.05}$ to effect the vertical registration of
cluster sequences for estimating relative ages, while Chaboyer
\etal\ (1996$a$) advocated the use of the subgiant-branch point
$V^-_{+0.05}$.  The advantages of estimating a fiducial point on the
subgiant branch are, first, that the sequence is more nearly horizontal
than the main sequence, and second, that the photometry for individual
stars is likely to be more precise.  A fiducial point on the main sequence
has the advantage that it can be estimated from a much larger number of
stars.  Furthermore, the shape of the upper main sequence is ``simpler,''
both in the sense that the high-order derivatives are smaller and in the
sense that the overall shape of the upper main sequence is less sensitive
to age, abundance, and uncertain details of stellar interiors than the
morphology of the subgiant branch (\eg, VBS Figs.~2 and 3).  However valid
these distinctions may be in general, with the present data sets both the
subgiant-branch and main-sequence fiducial points lead to much the same
relative shifts, as Table~3 shows, and hence ultimately to the same
astrophysical conclusions.  In fact, since the whole shape of the turnoff
region has been fitted as a fixed unit, when the various fiducial points
have been determined in this way all are determined to essentially the
same level of precision, and it makes little difference whether subsequent
analysis depends upon the turnoff magnitude or upon either of the
subiant-branch or main-sequence fiducial points.

We note that the isochrone fits imply turnoff colors approximately
0.007~mag bluer than the bluest normal points given in Table~2; possibly
this is because the nose of the turnoff is more pointed in the models than
in reality, but it is also possible that it is a consequence of a
different level of effectiveness in rejecting binaries and uncertain data
in the two schemes.  In either case, since it appears to be a constant
effect (at a level of $\pm\,0.003\,$mag or so), it is unimportant for
differential comparisons among clusters provided care is taken to base
those comparisons on one type of analysis at a time.  Apart from this
minor difference, we see generally excellent agreement between the
traditionally defined normal points and the somewhat more novel isochrone
fits, indicating that the latter method does provide a robust, impersonal,
and repeatable method for establishing the position and morphology
of a cluster locus from unbinned data.

\subsection{Intercomparison of the Outer Halo Clusters}

The first-order variations of our present isochrones with changes of age,
overall metal abundance, and $\alpha$-element enhancement are illustrated
in Figures~12--14\margin{Figs.~12--14}, where we have normalized a
selection of isochrones according to the precepts of VBS: the isochrones
are translated horizontally to match their turnoff colors (vertical
short-dashed line), and then they are shifted vertically to register the
point on the upper main sequence that lies exactly 0.05$\,$mag redder than
the turnoff (cross).  The variation of isochrone morphology with a change
in one of the assumed input parameters is then most easily perceived as a
change in the position of the giant branch.  For subsequent comparison
with our cluster data, we have parameterized these morphology variations by
the color difference between the turnoff and the point on the giant branch
(horizontal long-dashed line) that is precisely 3.0$\,$mag brighter in $V$
than the upper-main-sequence fiducial registration point (cross); this
corresponds to a magnitude level 2.19 or 2.20$\,$mag brighter than the
turnoff, as we have defined it above.  We will represent this color
difference by $(\vmi)_{GB} - (\vmi)_{TO}$, or $\Delta(\vmi)$.  Clearly,
when isochrones are compared to cluster data or cluster sequences are
compared to one another after registration in this fashion, the
differential interpretations that one may then make are independent of
cluster distance and reddening and of zero-point errors in both the
photometry and the theoretical isochrones.  We find that the change in
giant-branch color associated with a change in presumed age $\tau$ from 8
to 16~Gyr is $\delta[\Delta(\vmi)] = -0.115\,$mag, or
$\delta[\Delta(\vmi)]/\delta\log\tau = -0.382\,$mag/dex, which --- in
first-order expansion about an age of 12$\,$Gyr --- corresponds to about
--0.014$\,$mag/Gyr.  Similarly, to first order $\Delta(\vmi)$ decreases by
about 0.04$\,$mag for a 1$\,$dex increase in either [Fe/H] or
[$\alpha$/Fe].

The true cluster abundances [Fe/H] and [$\alpha$/Fe] are therefore
significant for interpreting these data, and we discuss current estimates
in detail in the Appendix.  The literature values are not definitive, but
they suggest that the three clusters are quite similar in [Fe/H], but with
Pal~3 likely to be the most metal-poor, Pal~4 the most metal-rich, and
Eridanus nearly as metal-rich as (and perhaps indistinguishable from)
Pal~4.  In Figure~15 \margin{Fig.~15} we compare the normal points of the
three clusters after dereddening and registration of their HB levels
(details to be discussed below).  The reddening values $E(\bmv)$ are taken
from Harris (1996), ``Catalogue of Milky Way Globular Cluster Parameters''
(see http://www.physics.mcmaster.ca/Globular.html, revision of
1997~May~15), and in our present analysis we have taken $E(\vmi) = 1.3
E(\bmv)$.  (This value comes from interpolation within Table~3 of
Cardelli, Clayton \& Mathis 1989, taking $V \sim 555\,$nm and $I \sim
814\,$nm; recall that the reddening actually occurs in the instrumental
magnitudes and not in the standard ones.  The reddening values estimated
for these five clusters are so small and so similar that our results are
quite insensitive to the value assumed for this ratio.)  For purposes of
this illustration only, we have assumed that all three clusters have an HB
luminosity $M_V = +0.70$ at $(\vmi)_\circ = 0.60$ (a color close to that
of the turnoffs).  This photometric comparison supports the conclusion
that the clusters are indeed similar in [Fe/H] and that the relative
metallicity ranking from the literature is probably correct:  Pal~3 has
the steepest and bluest RGB, while Eridanus and Pal~4 have RGBs with
slightly redder colors and flatter slopes that are virtually identical to
one another.  In addition, the HB of Pal~3 extends blueward into the
RR~Lyrae instability strip, whereas the HBs for both Eridanus and Pal~4
are simply red stubs with extremely small ranges in color, which --- all
other things being equal --- also tends to corroborate the inferred
metallicity ranking.

The near-coincidence of the main-sequence turnoff (``MSTO'') and subgiant
regions among all three clusters immediately suggests that they have
similar ages. Figure~16 \margin{Fig.~16} shows the three cluster fiducials
after registration via the VBS method. After such registration, any age
differences should be manifest as offsets in the cluster RGBs (again, all
other things --- such as composition --- being equal).  None is obvious in
this diagram.  However, we can perform a more quantitative test:
Figure~17 \margin{Fig.~17} shows the color-magnitude diagrams of Pal~3,
Pal~4, and Eridanus with the three clusters registered to match their
apparent turnoff colors and $V^+_{+0.05}$ fiducial magnitudes derived from
the isochrone fits (use of the fitted turnoff magnitudes or the
$V^-_{+0.05}$ fiducial points in effecting the vertical registration would
have made no material difference to the comparison).  To eliminate some
poorer measurements, here we consider only those stars with photometric
uncertainties $\sigma(\vmi) < 0.05\,$mag, where the standard error
assigned to a given star represents a compromise between that derived from
the expected Poisson and readout noise and that indicated by the actual
frame-to-frame repeatability for that particular star; 13\% of the
measured stars were rejected by this criterion.  After registration of the
data, we fit a simple parabola to the color of the giant-branch stars as a
function of visual magnitude for the three clusters taken together, over
the range $-4.1 < V$--$V^+_{+0.05} < -2.0\,$mag, $\left|\delta(\vmi)\right| <
0.07\,$mag, where $\delta(\vmi)$ is the horizontal distance of a given
point from the best-fitting parabola.  This region is denoted by the
curve-sided box in Fig.~17.  Then the net offset of the data for each
individual cluster from the fitted parabola was determined, as listed in
Table~4\margin{Table~4}:  the table gives (column~2) the mean differential
offset for the giant branch of each cluster, (column~3) the median offset
among the giants for each cluster, (column~4) the standard deviation about
the mean horizontal offset for the each cluster, (column~5)~the number of
stars contained in the box, and (column~6)~the median value of
$\sigma(\vmi)$ among the stars in the box.  We note that the perceived
standard deviation is quite a bit larger than our estimates of the
photometric errors for the individual stars.  There can be several reasons
for this:  (a)~we may have underestimated our photometric errors;
(b)~there could be actual scatter among the giants in each cluster;
(c)~the giant branch might not be a perfect parabola over this range of
magnitude; and (d)~the derived standard deviation could be dominated by
the random errors of stars with individual $\sigma(\vmi)$ values larger
than the median value.

Using the isochrones described above, we found that at [Fe/H]$\approx\,
-1.41$ and age $\approx12\,$Gyr, the relationship between RGB color offset
and age is $\delta[\Delta(\vmi)]/\delta \tau = -0.014$ mag/Gyr. If we take
the relative giant-branch offset of Pal~3 and Pal~4 to be of order $0.000
\pm 0.003\,$mag, then we would argue that these clusters are coeval to a
level of about 0.2$\,$Gyr {\it if the difference in their abundances can
be neglected\/} (although we will argue below that the abundance
differences should {\it not\/} be neglected).  On the other hand, the
Eridanus giant branch appears to be offset from the giant branches of the
two Palomar clusters by some $0.01 \pm 0.004\,$mag, which would suggest
that Eridanus is younger than the other two clusters by some $0.7 \pm
0.3$~Gyr under the same set of assumptions.

It is also quite apparent in Fig.~17 that by aligning the three cluster
sequences at the main-sequence fiducial magnitude $V^+_{+0.05}$, we have
also brought the three clusters' HBs into excellent coincidence.  The
small number of HB stars in each cluster perhaps makes it dangerous to try
to be overly quantitative (note in particular that the two double circles
on the HB bluer than $(\vmi)-(\vmi)_{TO} = 0.12$ represent mean values for
two of the Pal~3 RR~Lyrae variables; the other five were excluded by the
$\sigma(\vmi) < 0.05$ criterion), we feel that we may conservatively
estimate that the \dv\ values for the three clusters do not differ by more
than 0.05$\,$mag or so (we present actual measured values in Table~6
below).  As our isochrones suggest that at [Fe/H]$\approx -1.41$ and age
$\approx12\,$Gyr the turnoff fades by some +0.085$\,$mag/Gyr in $V$, this
would suggest an upper limit of some 0.6$\,$Gyr for the total age spread
among these clusters if their abundances are similar.  On the other hand,
it is commonly believed that at a fixed age the $V$-band luminosity of the
HB decreases by an amount of order 0.2$\,$mag for every 1$\,$dex increase
in the overall heavy-element abundance (\eg, Dorman 1993; Chaboyer,
Demarque \& Sarajedini 1996$b$; Gratton \etal\ 1997; our theoretical
models are consistent with this: VandenBerg \etal, in preparation).  If
this is true, and if Pal~3 is of order 0.2$\,$dex more metal poor than
Pal~4 and Eridanus, then its HB should be of order 0.04$\,$mag more
luminous than in those other clusters.  As we will discuss below, we
believe that an offset of this order of magnitude is within the measuring
error of the HB magnitude.

\subsection{Comparison with the Near-Halo Clusters \m5 and \m3}

With the CMD data for these three clusters we are in a position to begin
answering a two-decades-old question: are the outer halo clusters with
anomalously red HB morphologies truly younger than their counterparts in
the nearby halo? Even with these data from \hst, the answer to this
question will depend on the estimates of the abundances of the outer halo
clusters as well as for our representatives of the nearby halo, \m3 and
\m5.  (\m5 is currently inside the 8$\,$kpc boundary that nominally
separates the inner halo from the intermediate halo, but its large proper
motion --- Cudworth \& Hanson 1993 --- indicates that this situation is
only temporary.) From the arguments presented in the Appendix, we believe
that it is adequate at present to consider that Pal~4, Eridanus, and \m5
have essentially the same metallicity, while the metal abundances of Pal~3
and \m3 are similar to each other and perceptibly lower than those of the
other three clusters.

While the principal sequences of the intermediate-halo clusters \m3 and
\m5 can be compared to those of the outer-halo clusters by the same
techniques as we have just applied to the comparison of the outer-halo
clusters among themselves, it is best to remember that there are some
underlying differences.  The comparisons among the outer-halo clusters are
direct:  the observations were made with the same telescope (\hst\/) and
camera (WFPC2), and were subjected to an identical analysis and
calibration procedure.  \m3 and \m5, in contrast, were each observed with
several different ground-based telescopes and on numerous occasions.
Still, in many cases both clusters were observed on the same nights, and
all observations have been referred to a common photometric system with
the utmost rigor currently possible.  However unlikely, there remains the
possibility of some undetected difference between the \hst\ and
ground-based photometric systems that will render a comparison of the
nearby halo clusters to the outer-halo ones less robust than comparisons
that are strictly internal to either data set.  Uncertainties in the {\it
zero points\/} of the color and magnitude scales are not of immediate
concern; what might cause a significant problem would be uncorrected {\it
color dependences\/} in the transformations from the instrumental to the
standard magnitude systems, or a nonlinearity in the magnitude scale of
either the space- or ground-based photometry.  We stress that at present
we have no evidence that such problems exist.  Nevertheless, it would be
wise to retain an extra quantum of skepticism in the backs of our minds as
we interpret the comparison of the \hst\ data to the ground-based
observations.

We have repeated the determination of the relative giant-branch shifts for
the five clusters just as in the previous section.  This time we combined
the rectified ($V,\vmi)$ data for all five clusters and refit the mean
giant-branch parabola using the same constraints as before:  $-4.1 <
V$--$V^+_{+0.05} < -2.0$, $\left|\,\delta(\vmi)\right| < 0.07$.  We were a
little more stringent in our selection of stars for \m3 and \m5, retaining
only those stars with individual color errors $\sigma(\vmi) < 0.03\,$mag.
After the parabola had been fit, we determined the net offsets and scatter
for each of the individual clusters relative to the mean giant branch of
all of them as before; the results are listed in Table~5\margin{Table~5}.
The implications are that, if Pal~4, Eridanus, and \m5 all have the same
composition --- and we stress again that this means both [Fe/H] and 
[$\alpha$/Fe] --- then Pal~4 and Eridanus are younger than \m5 by some 1.4
and 2.1$\,$Gyr, respectively.  The difference between Pal~3 and \m3 ---
again, assuming that these two clusters have the same chemical abundances
--- is similar:  the net relative giant branch offset would imply that
Pal~3 is younger than \m3 by some 1.7$\,$Gyr.

The alternative technique of determining the relative ages of clusters
from the magnitude differences between their HBs and their main-sequence
turnoffs requires a robust, consistent measurement of their HB magnitudes
as well as of their turnoffs.  The upper-main-sequence and subgiant-branch
fiducial points given in Table~3 serve the latter purpose; now it remains
to try to estimate the fiducial HB magnitudes with similar rigor.  We have
computed zero-age horizontal branches (ZAHBs) consistent with the
predictions of these main-sequence and giant-branch evolutionary tracks.
We have employed them much as we used the isochrones in \S3, simply as
drafting splines of approximately the right shape to match the shape of an
actual cluster HB.  For comparison with Pal~3 and \m3, we have adopted
theoretical ZAHB sequences with [Fe/H] = --1.54, [$\alpha$/Fe] = +0.3, and
$Y = 0.2362$; Pal~4, Eridanus, and \m5 were compared to theoretical curves
with [Fe/H] = --1.41, [$\alpha$/Fe] = +0.3, and $Y = 0.2366$.  Unlike the
case with the turnoff isochrones, in this case we have allowed no freedom
for horizontal translation of the theoretical curves, but have adopted
their predicted colors and estimated reddening values verbatim, and have
performed vertical shifts only in matching the theoretical ZAHBs to the
lower envelope of the dereddened cluster data.  Figure~18\margin{Fig.~18}
shows these comparisons.  The apparent visual magnitude of the fitted ZAHB
was read out at an intrinsic color of $(\vmi)_\circ = 0.60$, and the
resulting $V_{HB}$ values are listed in Table~6\margin{Table~6}.  The
external accuracy of these derived HB magnitudes is clearly
dependent upon the requirement that the theoretical models {\it
accurately\/} reproduce the luminosity variation of the HB
with color, both at the red end (for the three outer-halo clusters), and
at the blue end (for the two nearby clusters):  if these curves are
seriously wrong in detail, errors in the vertical placement of the ZAHB of
several hundredths of a magnitude could easily result.

The precison of the HB magnitudes is also plainly jeopardized by the small
number of HB stars in each cluster.  For the three distant systems, there
is little more that can be done; these diagrams contain most of the HB
stars that Nature has provided.  For \m5, at least, we have a convenient
check.  Sandquist \etal\ (1996) have published $(V,I)$ photometry for a
large sample of stars in this cluster.  Johnson \& Bolte (1998) find some
modest errors in the zero points and possibly the scale of the Sandquist
\etal\ magnitudes and, based on our own analysis, we agree.  However,
considering only those stars common to the Sandquist \etal\ sample and our
current data set with $14.5 < V < 16.5$ --- 60 stars in all --- we find
net differences of $\Delta V = +0.009 \pm 0.0029\,$mag (standard error of
the mean, $\pm 0.022$ standard deviation of one difference) and $\Delta I
= -0.006 \pm 0.0026\,$mag (s.e.m., $\pm 0.020$, s.d.) in the sense
(present -- Sandquist \etal).  The data of Sandquist \etal\ for stars with
$\sigma(\vmi) < 0.03\,$mag and lying more than 96\sec\ from the center of
\m5 are plotted, after correction for these offsets, in
Figure~19\margin{Fig.~19} along with the same fitted ZAHB as before.
[Note that here $\mmm_V$ does {\it not\/} necessarily mean $-5 + 5\log
d\hbox{(\rm pc)} + A_V$; rather it means
m$_V\hbox{(observed)}-$M$_V\hbox{(predicted)}$.]  It seems that in this
case, at least, our derived HB magnitude is consistent with the larger
data set.

The last column of Table~6 lists the values of $V_{TO} - V_{HB}$ that
result from the difference between these HB magnitudes and the turnoff
magnitudes derived from the isochrone fits.  We see that for the three
outer-halo clusters, the magnitude differences between the HB and the
turnoff are  nearly indistinguishable.  Likewise, the difference between
\m3 and \m5 is infinitesimal, but there is a difference of approximately
0.15$\,$mag in the turnoff-to-HB magnitude difference (or the
subgiant-branch-to-HB difference) between the nearby and remote clusters.
In the approximation where all clusters have effectively the same chemical
abundances, this would imply that the three outer-halo clusters are all
roughly 1.5 -- 2$\,$Gyr younger than the two near-halo clusters.

To further quantify this conclusion, it is necessary to look more closely
at the predictions of the current set of isochrones, in particular at the
dependence of $\Delta(\vmi)$ and $\Delta V$ on chemical abundance in
comparison to their dependence on age.  The solid curve in
Figure~20\margin{Fig.~20} shows the predictions of VandenBerg's isochrones
for the evolution of the turnoff-to-giant-branch color difference
(horizontal axis) and the turnoff-to-HB magnitude difference (vertical
axis) as functions of age, from 8 to 18$\,$Gyr, for abundances of [Fe/H] =
--1.41 and [$\alpha$/Fe] = +0.3.  The horizontal tick marks intersect the
curve at the model predictions for incremental age increases of 2$\,$Gyr.
The arrow illustrates the effect on the models produced by an assumed
increase of +0.2~dex in either [Fe/H] or [O/Fe], according to this
particular set of isochrones.  There is considerable uncertainty in the
horizontal placement of the solid curve in this diagram, as the predicted
colors are dependent both upon uncertain details of convection theory and
upon the assumed conversion from theoretical effective temperatures to
observational colors.  The slope of the curve could also be somewhat in
error if the assumed dependence of convection or the temperature-to-color
conversion on temperature itself is in error in a seriously non-linear
fashion.  Similarly, there could be some uncertainty in the absolute
vertical positions of the tickmarks if the predicted HB luminosities are
in error, or if the conversion of bolometric magnitudes to $V$-band
magnitudes is incorrect for either the turnoff or HB stars.  Therefore we
particularly wish to de-emphasize any implication that we are measuring
absolute ages for any of these clusters, at least until some of the
remaining uncertainties in the model physics have been resolved.  However,
the size of the intervals between the vertical positions of the tick
marks, representing increments of order 15--20\% in age, should be
relatively secure.  The locations in this plane of the three outer-halo
clusters are represented by open symbols, a triangle for Pal~3, a square
for Pal~4, and a circle for Eridanus, while the two nearby clusters are
indicated by filled symbols: a triangle for \m3 and a pentagon for \m5.
Representative one-sigma error bars of $\pm0.005\,$mag in $\Delta(\vmi)$
and $\pm0.05\,$mag in $\Delta V$ --- due primarily to the uncertain
placment of the HB lower envelope --- are illustrated in the bottom right
corner.

A consistent interpretation of the two age indicators can be made for the
five clusters.  {\it If we assume that the theoretical isochrones employed
here correctly predict the dependence of $\Delta V$ and $\Delta(\vmi)$ on
age and chemical abundance\/}, then:
\begin{enumerate}
\item Those same models {\it
slightly\/} (\ie, by about 0.017$\,$mag) overestimate the color difference
between the giant branch and the turnoff, and the solid curve should be
shifted as a body to the left.

\item The data are consistent with an age difference of some 2$\,$Gyr
between \m3 and Pal~3 (Pal~3 younger) if their abundances are the same.
Similarly, Pal~4 is of order 1.5$\,$Gyr younger than \m5 if their
abundances are identical.

\item The difference between the CMDs of Pal~4 and Eridanus could be
due to 
\begin{description}
\item[\rm a.] an age difference of somewhat less than 1$\,$Gyr (Eridanus
younger) if their abundances are the same; or 

\item[\rm b.] an abundance difference of some 0.2$\,$dex with Eridanus being
the more metal poor, in which case the age difference would be small.
\end {description}
The available spectroscopic data suggest that Eridanus {\it may\/} be some
0.14$\,$dex more metal-poor than Pal~4, so the evidence for an age
difference between the two is tenuous.

\item Even though the point for Pal~3 coincides with that for Pal~4 in
this diagram, it is likely that Pal~3 is actually $\sim\,$1$\,$Gyr older
than Pal~4, because its point will have been displaced upward and to the
right in this diagram as a result of its lower metal abundance.

\item Similarly, \m5 is almost certainly more metal-rich than \m3, so by
the same argument as in point 4 above,  \m5 must also be younger than \m3,
probably by of order 1$\,$Gyr or somewhat over.  \end{enumerate}

\section{Discussion}

The two relative age indicators, $\Delta(\vmi)$ and $\Delta V$, suggest a
single self-consistent interpretation, namely that all three outer halo
clusters are younger than their near-halo counterparts by 1.5 to 2$\,$Gyr,
if we have correctly estimated the clusters' chemical abundances.
Sarajedini's (1997) result for Pal~14 places this fourth outer-halo
cluster in the same age ballpark (with, of course, the same caveats
regarding uncertainties in the chemical abundance ratios and poorly
understood processes in stellar interiors).  Based on LDZ's HB star
models, if the red HB of Pal~3 is due to an age difference between this
cluster and the average [Fe/H]$\sim -1.5$ cluster in the near halo, Pal~3
is $\sim$$\,2\,$Gyr younger than its near-halo counterparts. For Pal~4 and
Eridanus, with HB types $(B-R)/(B+V+R) = -1$, Lee's method can assign only a
lower limit of $\geq 2\,$Gyr to the age difference, since the models for
all ages younger than this imply wholly red HBs.  These differential ages
based on HB morphologies are in the same sense and of approximately the
same size as those implied by our turnoff and subgiant comparisons.

Another angle from which to view this result is to ask, ``What other
parameters of the outer-halo clusters would have to be different if they
are to have the {\it same\/} ages as the near-halo clusters?''  We
conclude that their age differences would be reduced to less than
$\sim\,$1 Gyr (below which we would find it difficult to argue
convincingly for genuine age differences) if we have overestimated the
heavy-element abundances for the outer-halo objects --- either [Fe/H] or
[$\alpha$/Fe], or a combination of the two --- by 0.3 dex or more.  We
cannot rule out this possibility or, for that matter, less frequently
discussed ones such as differences in helium abundance, mass loss rates on
the giant branch, or stellar rotation.  As we discuss in the Appendix, the
[Fe/H] determinations for the outer-halo clusters are considerably weaker
than those of the near-halo ones, and we have no quantitative information
at all on their [$\alpha$/Fe] values.  It is intriguing in this context
that recent spectroscopic observations by Brown, Wallerstein \& Zucker
(1997) indicate that the young halo globular clusters Ruprecht~106 and
Palomar~12 have relative oxygen and general $\alpha$-element abundances
[O/Fe] and [$\alpha$/Fe] $\sim 0.0$ to +0.1, in contrast to the value +0.3
to +0.4 observed in nearby, ``normal'' halo objects, although in the case
of oxygen it is unknown whether the abundance ratio in Rup~106 and Pal~12
is primordial or has been altered by mixing.  If Pal~3, Pal~4, and
Eridanus were to have primordial abundances [$\alpha$/Fe] $\sim 0.0$, then
by basing our metallicity on measurements of Ca calibrated via nearby
clusters with [$\alpha$/Fe]$> 0$ we have assigned the wrong [Fe/H]. The
sense of this error is such that it would reduce the inferred age
difference between them and the near-halo clusters.  We therefore believe
it is of the utmost importance to obtain high-resolution spectroscopic
data for the outer-halo clusters.

We have been scheduled observing time on \hst\ to obtain color-magnitude
diagrams for the last two remaining globular clusters beyond 50$\,$kpc
from the Galactic center, Palomar~14 and AM$\,$1.  At the same time, other
research teams are using \hst\ to pursue the study of the globular
clusters near the Galactic center, and the dwarf galaxies that share the
outer halo with the Palomar-like clusters, while still others are
improving the available data for the near-halo objects.  Thus, this is
probably not the time to draw definitive conclusions regarding the early
evolution of the Galactic spheroid.  We can nevertheless draw some
provisional conclusions that may be tested and rejected or refined by
future work.  As always, these conclusions are subject to the provisos
that we have correctly estimated the chemical abundances of the objects
under discussion and that no important physics has been omitted from the
{\it differential\/} predictions of stellar evolution; simple errors in
distance scale, convection theory, opacities, elementary particle physics,
and anything else can be neglected provided they affect all globular
clusters comparably, and provided we restrict ourselves to investigating
the order and relative time intervals over which things happened, as
opposed to the exact moment that any one event took place.

These results make a good case for age as a second parameter: the mean age
of the outer-halo clusters with anomalously red HBs appears to be lower by
$\sim 1 - 2$ Gyr than that of the nearby, ``normal'' clusters.  Although
the original motivation for many programs investigating ages for GGC has
been to decide the mix of ELS and SZ that is required to match the cluster
age distribution, it is clear that the GGC age distribution alone does not
lead directly to a definitive formation scenario for the Galaxy.  A range
in GGC ages ($>$ a few  Gyr) is clearly inconsistent with the most naive
version of a monolithic collapse model, but a general large-scale lumpy
collapse accompanied or followed by various degrees of infall or accretion
are allowed.  Hierarchical formation models, like SZ, can more easily
accomodate a range of ages than ELS, but they do not make {\it specific\/}
predictions of the amount of time required to assemble the Galactic Halo
and they do not predict an age distribution for GGC.  Observationally, the
Sagittarius dwarf spheroidal galaxy provides a striking example of a
``late'' accretion event that will add field stars and globular clusters
to the Galaxy, yet not significantly alter the age dispersion of the GGC
system. Still, it would be premature to conclude on the basis of the
Sagittarius example that accretion was necessarily the sole, or even the
dominant, mode of halo formation.  Based on high-resolution simulations of
the formation of galaxies after recombination (\eg, Sommerville \&
Primack 1998), there is a large overlap in the progenitor collapse times
of objects that become galaxies of different masses.  Objects that
eventually become LMC-like galaxies can first start to form stars before,
at the same time or after Milky Way-like objects.  The age distribution is
a crucial piece of the picture, but the final story will depend on other
information such as detailed abundance patterns and kinematics for GGC
halo populations (\eg, Carney \etal\ 1997; Zinn 1993).

In contrast to theories that attempt to describe the hierarchical
formation process of the Galaxy as a whole, the Harris \& Pudritz (1994)
formation model for the globular clusters themselves does make specific
quantitative predictions for the amount of time required to assemble these
systems within the outer Galactic halo.  Far out in the newly forming
Galaxy, the gas clouds (supergiant molecular clouds or SGMCs) in which the
protoclusters are presumed to have been built would have had
characteristically larger linear sizes, to match the lower ambient
pressure that was found there, in comparison to the denser conditions in
the rapidly collapsing Galactic core.  The timespan over which
protoclusters can form scales directly with the internal free-fall time of
the SGMC; thus, the cluster formation timespread will increase as the
surrounding pressure drops.  If the SGMC lifetime in the mid-halo region
was typically $\sim 0.5$ Gyr (see Harris \& Pudritz equation 5.9 and
related discussion), then it could have been $\gtsim 2$ Gyr in the
outermost regions.

Finally, as already noted by van den Bergh (1998), the GGC which have been
shown to be young are all at the low-luminosity end of the $M_V$
distribution and are all beyond $R_{\rm GC} = 15$ kpc.  The massive
clusters beyond this Galactocentric radius do appear to be co-eval with
the oldest near-halo clusters.  Van den Bergh interprets this to indicate
that the mass of newly-forming clusters decreased with time in the outer
halo.  Our present data agree with this interpretation. It will be very
interesting to determine reliable ages for the remaining clusters at large
$R_{\rm GC}$ (all relatively low-L objects) and to pursue similar studies
of the low-L GGC closer to the center of the Galaxy.

\acknowledgments 

Support for this work was provided by NASA to M.B., H.E.B., and R.A.B.
through STScI grants GO-06106.01-94A, .02-94A, and .03-94A from the Space
Telescope Science Institute, which is operated by the Association of
Universities for Research in Astronomy, Incorporated, under NASA contract
NAS5-26555.  Financial support from the Natural Sciences and Engineering
Research Council of Canada, through research grants to W.E.H., D.A.V.,
G.G.F., and H.B.R. is also gratefully acknowledged.

\clearpage

\appendix

\section{[Fe/H] and [$\alpha/H$] Determinations from the Literature}

Extracting reliable relative {\it or\/} absolute age information from CMDs
depends upon having accurate relative {\it and\/} absolute values for the
chemical compositions of their constituent stars. In particular, the
morphology of the HB is sensitive to (often unmeasurably) small
differences in many parameters, among them age, [Fe/H], [CNO/Fe],
[$\alpha$/Fe], $Y$, mass, core rotation, and so on. Thus, constraining the
role played by age differences in producing a range of HB morphologies
among clusters requires understanding the roles played by all other
variables.  A special emphasis emerges on understanding possible chemical
differences which --- unlike, say, interior rotation --- we can, in
principle, directly measure.  (However, note recent papers by Rutledge,
Hesser \& Stetson 1997 and King \etal\ 1998, which demonstrate that there
remains considerable doubt about the significance of the chemical
abundances that can be inferred from existing spectroscopic observations.)
Similarly, differential comparisons among remote and nearby halo objects
offer a powerful tool to constrain possible age gradients as a function of
Galactocentric radius, provided that we can identify near-halo clusters
with compositions comparable to those of the outer-halo objects.  In this
section, we review information from the literature on our outer halo
clusters and the comparison clusters we have chosen for them.

\subsection{General remarks}

It is a risky undertaking to combine abundances from different literature
studies because there is not a general agreement on the overall abundance
scale.  The most widely used general metallicity scale is that of Zinn \&
West (1984; hereinafter ZW84), which is ultimately based upon photometric,
spectrophotometric, and photographic-spectrum abundance determinations
from the 1970's. Although some recent studies have called into question
the zero point and scaling of absolute metallicity values from ZW84 (\eg,
Carretta and Gratton 1997), it appears to provide generally reliable
relative {\it rankings\/} of cluster metallicity (\eg, Rutledge \etal\ 
1997).  Of greatest relevance to our present study, we note that
in the range $-1.0 < {\rm [Fe/H]} < -1.9,$ the Zinn-West values are
systematically lower by 0.23 dex than the Carretta-Gratton values.
Similar trends are seen when comparing Rutledge \etal's large, homogeneous
survey of \ion{Ca}{2} triplet measures to the two scales.  We are
fortunate that Pal~3, Pal~4 and Eridanus have all been the subject of
abundance studies which allow them to be differentially compared to our
two ``standard'' clusters \m3 and \m5 (see details below); to some
extent, it is this relative ranking by metallicity that is most important
for determining the range of ages among the clusters.

Because we will put high weight on [Fe/H] determinations for the
outer-halo clusters that have been inferred from the lines of Ca in the
near-IR, we need to be concerned not only with [Fe/H] but also with the
size of the cluster-to-cluster variations in [Ca/Fe] or more generally
[$\alpha$/Fe].  In recent years considerable observational evidence has
accumulated that near-halo clusters and field stars generally display
roughly constant ratios of $\alpha$ to iron-peak elements, with the former
being more abundant than the latter by several tenths of a dex for
[Fe/H]$< -1.0$; the $\alpha$-to-iron ratio approaches the solar value as
[Fe/H] increases from --1.0 to 0.0, at least among the field stars
surveyed.  In a review of elemental globular cluster abundances based
again on modern high-dispersion studies, Carney (1996) concludes that
there is no variation of [$\alpha$/Fe] as a function of metallicity for 14
nearby clusters, all with [Fe/H]$\,\ltsim\,$--0.6 to --0.7. He further
summarizes the evidence for an $\alpha$-element enhancement relative to
iron of [$\alpha$/Fe] $\sim\,$+0.28~dex for these objects.

The great distances of the most remote clusters make fine abundance
analyses challenging, and there is so far little direct evidence bearing
on [$\alpha$/Fe] trends in the outer halo.  Indeed, there are only a few
such studies for clusters at intermediate $R_{\rm GC}$. Brown, Wallerstein
\& Zucker (1997) have found that two giant stars in each of Ruprecht~106
and Palomar~12 exhibit solar ratios of the $\alpha$ elements
(specifically, Mg, Si, Ca, and Ti), yet the clusters' overall
heavy-element abundances ([Fe/H]=$-1.45\pm 0.10$ and $-1.0\pm0.10$,
respectively) fall in the regime where most halo objects exhibit enhanced
$\alpha$ elements.  Moreover, for Rup~106 they detected an oxygen anomaly,
in that its value of [O/Fe]$=0.0\pm0.13$ is about 3$\sigma$ below that
normally found for stars with [Fe/H] $\sim -1.5$.  Could such anomalies
also occur in other outer-halo clusters?  We have to be cautious in
interpreting the latter measurement because [O/Fe] has been shown in
several clusters to decrease with increasing RGB luminosity, suggesting
that CNO-processed material is being dredged into the star's atmospheres.
The low [O/Fe] for the brightest Rup~106 stars may therefore not reflect
the initial oxygen abundance of the stars.  Based on six giants in
\ngc{7006} which, at $R_{\rm GC} = 38\,$kpc, is half of the way out to the
far halo, Kraft \etal\ (1998) find [Ca/Fe]$\sim +0.2$ --- consistent with
the general trend of the near-halo clusters but also, at the other extreme
of the error bar, possibly consistent with less $\alpha$-element
enhancement than usual in the near halo.  With its very large space
velocity (Cudworth \& Hanson 1993), \m5 might itself be considered an
outer halo denizen that shows ``normal'' [$\alpha$/Fe].

For Population II field stars, there have been additional indications that
[$\alpha$/Fe] may be solar or subsolar for some objects with very large
space velocities and therefore large inferred apogalactic distances (\eg,
Carney \etal\ 1997; King 1997). It is too soon to draw any general
conclusions about systematic trends in [$\alpha$/Fe] with $R_{\rm GC}$.
In what follows we must keep in mind that these remote objects may have
[$\alpha$/Fe] ratios that are less enhanced than those of our fiducial
clusters \m3 and \m5, and from the other calibrating clusters for the
metallicity scales.

\subsection{Eridanus}

Of the three outer-halo clusters, only Eridanus was included in the
original ZW84 compilation; they assigned it [Fe/H]$=-1.4$, the same as \m5
and $\sim\,$0.25$\,$dex more metal-rich than \m3.  Armandroff \& Da Costa
(1991) obtained \ion{Ca}{2} triplet spectra for two Eridanus giants along
with numerous giants in six nearby well-studied clusters. Their reduced
calcium equivalent widths suggested that Eridanus falls about one-third of
the way from \ngc{6752} ([Fe/H]$=-1.55$) to \ngc{1851} ([Fe/H]$=-1.26$),
two of their ``standard'' clusters. Thus from the \ion{Ca}{2} measures
alone, they derived $-1.41\pm0.11$ on the ZW84 scale.  (The same placement
relative to \ngc{6752} and \ngc{1851}, interpreted through the Rutledge
\etal\ version of ZW84  and the Carretta-Gratton scales would be $-1.44$
and $-1.14$, respectively.  All three of these estimates are internally
consistent to within $\pm 0.1$ dex, given the 0.23 dex offset between ZW84
and CG97 mentioned above.)  Armandroff \& Da~Costa combined their
\ion{Ca}{2} abundance values with two earlier estimates ($-1.50\pm0.15$ by
Da~Costa \& Armandroff (1990), derived from the position of the Eridanus
giant branch in the (M$_I$, (V$-$I)$_\circ$) plane; and Da~Costa's (1985) value
of [Fe/H]$=-1.35\pm0.2$ determined from (\bmv)$_{\circ{\rm ,g}}$) to adopt a
mean $-1.42\pm0.08$ dex for the cluster (again on the ZW84 scale).

Earlier, Ortolani \& Gratton (1989) had analyzed low-signal-to-noise,
low-dispersion spectra of two Eridanus giants and four Pal~3 giants (see
below). They concluded that Eridanus falls between \ngc{3201} and
\ngc{4590} in metallicity and assigned it [Fe/H]$=-1.6\pm0.3$ on their
abundance scale; since ZW84 assigns metallicities of --1.6 and --2.1 to
\ngc{3201} and \ngc{4590}, respectively, Ortolani \& Gratton's spectra
might imply [Fe/H]$\,\sim\,$--1.8 on the ZW84 scale for Eridanus, but the
error bar is large.

\subsection{Pal~3}

Pal~3 and Pal~4 were included in the investigation of Armandroff, Da~Costa
\& Zinn (1992) in which abundances were measured from the \ion{Ca}{2}
triplet. Spectra for three Pal~3 giants were measured and the reduced
equivalent widths placed Pal~3 more metal-poor than \m5 and similar to
\m{13}. In this study, Pal~3 was also ranked more metal-poor than Pal~4.
On the ZW84 scale, Pal~3 was assigned [Fe/H]$=-1.57\pm 0.19$.

Pal~3 was also part of the Ortolani \& Gratton (1989) study mentioned
above. From low-dispersion spectra of four giants, Pal~3 was assigned
[Fe/H]$=-1.6\pm 0.3$ (the same as Eridanus from that study) although the
star-to-star scatter was large.  From their CMD for Pal~3, Ortolani \&
Gratton estimated $E(\bmv)=0.03\pm0.02$ and $(\bmv)_{\rm o,g} = 0.69\pm
0.04$, leading to [Fe/H]$=-1.88\pm0.15$.  They adopted $-1.7\pm0.2$ from
this combination of spectral and CMD indicators.

\subsection{Pal~4}

\ion{Ca}{2} triplet measures for three giants led  Armandroff \etal\ (1992) 
to conclude that [Fe/H]$=-1.28\pm0.20$ for Pal~4 on the ZW84 scale,
compared to earlier estimates from CMDs of [Fe/H]$=-1.7$ (Christian
\& Heasley 1986) and [Fe/H]$\sim-1$ (Reed \& Harris 1986). They further
note that their triplet metallicity determination would be
$-1.35\pm0.14$ were \ngc{6171} excluded from their calibration.
>From their Figure~4, Pal~4 appears to be clearly more metal-rich
than \m{13} and Pal~3 and comparable to or slightly more metal-rich 
than \m5.

\subsection{Halo Cluster Metallicity Summary}

The previous studies are in good general agreement in finding 
the Eridanus metallicity to be $\sim -1.4$ on the ZW84 scale.
To the extent that \ngc{6752} is similar in [Fe/H] to \m3 and \m{13},
Eridanus appears to be more metal-rich than \m3.
Based primarily on the Armandroff \etal\ \ion{Ca}{2} triplet work, we
conclude that Pal~3 is closer to \m3 and \m{13} in metallicity
than it is to \m5. Furthermore, by appealing to the differential
studies of Armandroff \etal\ (1992) and Ortolani \& Gratton (1989)
we suggest that Pal~3 is the most metal-poor of our three
clusters.

For purposes of comparison, our adopted [Fe/H] abundances for all five
clusters, {\it on the ZW84 scale}, are listed in Table~6.  The values for
\m3 and \m5 are taken from the compilation of Harris (1996); each of the
five is likely to be externally uncertain at the $\pm 0.15\,$dex level at
least (see Paper I).  Nevetheless, they indicate formally that there is no
significant difference between \m3 and Pal~3, or between \m5 and Pal~4.
On the strength of both the metallicity measurements and the CMD
comparisons shown above, we believe Eridanus to be most nearly similar to
\m5 as well.

\clearpage

\figcaption[fig1.ps]{Color-magnitude diagram for the outer-halo globular
cluster Pal~3, derived from our WFPC2 photometry with the \hst.  Only
stars with photometric standard errors $\sigma(\vmi) < 0.10\,$mag have
been plotted, and data from all four CCD fields have been combined.  The
circled cross at the blue end of the HB represents the mean
magnitude and color of 7 RR~Lyrae stars in the cluster.  \label{fig1}}

\figcaption[fig2.ps]{Color-magnitude diagram for Pal~4, as in Fig.~1. 
\label{fig2}}

\figcaption[fig3.ps]{Color-magnitude diagram for Eridanus, as in Fig.~1.
\label{fig3}}

\figcaption[fig4.ps]{Ground-based ($V,\vmi$) color-magnitude diagram for
the nearby globular cluster M3.  Only stars with photometric standard
errors $\sigma(\vmi) < 0.10\,$mag have been plotted.\label{fig4}}

\figcaption[fig5.ps]{Ground-based ($V,\vmi$) color-magnitude diagram for
the nearby globular cluster M5, as in Fig.~4.  
\label{fig5}}

\figcaption[fig6.ps]{Light-curve fragments for seven candidate RR~Lyrae
variables in the outer-halo globular cluster Pal~3, based upon
our \hst\ observations.  Open squares represent instantaneous
$I$-band magnitudes, and closed circles represent $V$-band magnitudes.
\label{fig6}}

\figcaption[fig7.ps]{The open squares are our fiducial points for the CMD
of Pal~3, as determined by mean colors after three iterations of
$\sigma$-clipping in bins of height 0.2$\,$mag in $V$.  The
solid curve is a robust maximum-likelihood fit of which isochrone that best
matches the observed CMD from among a wide variety of
isochrones that were tried.  For the HB level and brighter, the individual
HB and RGB stars are plotted (crosses).  \label{fig7}}

\figcaption[fig8.ps]{Fiducial points and isochrone fit for the
CMD of Pal~4, with symbols as described in Figure~7.  \label{fig8}}

\figcaption[fig9.ps]{Fiducial points and isochrone fit for the CMD of
Eridanus, with symbols as described in Figure~7.  \label{fig9}}

\figcaption[fig10.ps]{Fiducial points and isochrone fit for the CMD of
\m3, with symbols as described in Figure~7.  \label{fig10}}

\figcaption[fig11.ps]{Fiducial points and isochrone fit for the CMD of
\m5, with symbols as described in Figure~7.  \label{fig11}}

\figcaption[fig12.ps]{Theoretical isochrones for chemical-abundance
ratios [Fe/H] = --1.41 and [$\alpha$/Fe] = +0.3, and ages of
8 to 16$\,$Gyr in steps of 2$\,$Gyr, illustrating the predicted
dependence of CMD morphology on age, as described in the text.
\label{fig12}}

\figcaption[fig13.ps]{Theoretical isochrones for chemical-abundance ratios
[Fe/H] = --1.14, --1.41, and --1.61;  [$\alpha$/Fe] = +0.3; and an age of
12$\,$Gyr, illustrating the predicted dependence of CMD morphology on
relative iron abundance.  The isochrones have been shifted horizontally
and vertically as in Fig.~12.\label{fig13}}

\figcaption[fig14.ps]{Theoretical isochrones for [Fe/H] = --1.41;
[$\alpha$/Fe] = 0.0, +0.3, and +0.6; and an age of 12$\,$Gyr, illustrating
the predicted dependence of CMD morphology on relative $\alpha$-element
abundance.  The isochrones have been shifted horizontally and vertically
as in Fig.~12.\label{fig13}}

\figcaption[fig15.ps]{Fiducial points for the three outer-halo clusters
(Pal~3, Pal~4, and Eridanus), dereddened according to the values listed in
Table~6, and shifted vertically to match their red HBs (the
short horizontal line at $M_V = 0.70$ represents the RHB level of each
cluster).  The distance scale for $M_V$(HB) has been chosen arbitrarily
for comparison purposes. Note that the mean points for Pal~4 and
Eridanus are essentially indistinguishable; the giant branch of Pal~3
lies $\simeq 0.02$ magnitude to the blue of the GBs of the other two
clusters.  \label{fig15}}

\figcaption[fig16.ps]{The mean points for Pal~3, Pal~4, and Eridanus have
been re-registered as prescribed by VandenBerg, Bolte, \& Stetson (1990):
the CMDs have been shifted horizontally so that the MSTO colors coincide
(vertical line), and vertically so that the point on the main sequence
0.05 mag redder than the MSTO coincides (horizontal line).  The horizontal
shifts were taken from the isochrone fits (Table~3), which produce
estimated turnoff colors systematically 0.007$\,$mag bluer than
$\sigma$-clipped normal points given in Table~2; thus there appears to be
a slight offset of the turnoff points from $\Delta(\vmi) \equiv 0.00.$
This offset is nearly identical for the three clusters, so no material
harm is done to the differential analysis.  For a given composition, an
age difference appears as an offset in the RGB color.  Pal~4 and Eridanus
appear to be indistinguishable in age, while Pal~3 may be slightly older.
If the clusters do not have identical abundances, the interpretation of
this diagram becomes more complex, as described in the text.
\label{fig16}}

\figcaption[fig17.ps]{The data for Pal~3, Pal~4, and Eridanus have
been registered as in Fig.~16, except in this case the stars have
been plotted individually.  The curve-sided box represents the region
where the relative displacements of the three giant branches were
determined, as listed in Table~4.  Only stars with $\sigma(\vmi) <
0.05\,$mag have been plotted.  Symbol types are as in Fig.~16.
\label{fig17}}

\figcaption[fig18.ps]{Stars in the region of the red HB
have been plotted for (from top to bottom) Pal~3, Pal~4, Eridanus,
\m3, and \m5.  The photometry has been dereddened according to the
color excesses in Table~6, with the assumption that
$E(\vmi) = 1.3 E(\bmv)$.  The solid curves represent our eyeball fits
of theoretical ZAHB loci to the lower envelope of the horizontal-branch
stars for each cluster.\label{fig18}}

\figcaption[fig19.ps]{Horizontal-branch region of the CMD of \m5
based upon the $V,I$ data of Sandquist \etal\ (1996), after correcting
for minor zero-point shifts in the photometry as described in the
text.  The solid curve is the same ZAHB locus as in the bottom panel
of Fig.~18.  Here, $\mmm_V$ denotes the vertical shift required to
register theoretical luminosities $M_V$ with observed $V$-band
magnitudes; due to uncertainties in the modelling, this may not
accurately represent the actual geometric distance of the cluster.\label{fig19}}

\figcaption[fig20.ps]{The relationship between the giant-branch-to-turnoff
color difference (abscissa) and the horizontal-branch-to-turnoff magnitude
difference (ordinate), according to the present set of isochrones.  The
solid curve represents the predicted track followed by a globular cluster
with [Fe/H] = --1.41 and [$\alpha$/Fe] = +0.3 as it ages from 8$\,$Gyr
(top right) to 18$\,$Gyr (bottom left) according to the present set of
theoretical models.  The absolute placement of the curve and tick marks
are uncertain due to incomplete knowledge of the physics of stellar
interiors, but the size of the intervals, representing age differences of
2$\,$Gyr, or 15--20\%, should be correct in a relative sense.  The arrow
represents the amount by which the models would be shifted for an iron or
$\alpha$-element abundance 0.2$\,$dex larger.  The positions in this
diagram currently occupied by the outer-halo globular clustars are
represented by open symbols:  triangle (Pal~3), square (Pal~4), and circle
(Eridanus).  The near-halo clusters are represented by closed symbols:
triangle (\m3) and pentagon (\m5).  The cross in the lower right
represents our estimate of the random observational errors ($\pm 1\sigma$)
in the two coordinates.  \label{fig20}}

\clearpage
\begin{deluxetable}{lcrcrcc}
\tablecaption{Notional Periods for Candidate RR~Lyrae Variables}
\tablecolumns{7}
\tablewidth{0pt}
\tablehead{
\colhead{} & \colhead{} & \colhead{HJD--} & \colhead{} & \colhead{HJD--} \nl
\colhead{Star ID} & \colhead{extremum} & \colhead{2$\,$449$\,$800} & 
\colhead{extremum} & \colhead{2$\,$449$\,$800} & \colhead{$\Delta t$} & \colhead{$ n\times P$?}
}
\startdata
1-233 & $I$ minimum & 20.68~~~ & $V$ minimum & $\gtsim\,$22.49~~ & 1.81 & 3$\times$0.60 \nl
1-299 & $I$ minimum & 20.82~~~ & $V$ minimum & $\ltsim\,$22.01~~ & 1.19 & 2$\times$0.60 \nl
1-404 & $I$ minimum & 20.62~~~ & $V$ minimum & 22.42~~ & 1.80 & 3$\times$0.60 \nl
2-198 & $I$ minimum & 20.61~~~ & $V$ minimum & 22.35~~ & 1.74 & 3$\times$0.58 \nl
2-368 & $I$ maximum & 20.55~~~ & $V$ maximum & 22.36~~ & 1.81 & 3$\times$0.60 \nl
2-614 & $I$ minimum & 20.82~~~ & $V$ minimum & $\ltsim\,$22.01~~ & 1.19 & 2$\times$0.60 \nl
3-269 & $I$ maximum & 20.61~~~ & $V$ maximum & 22.42~~ & 1.81 & 3$\times$0.60 \nl
\enddata
\end{deluxetable}

\clearpage
\begin{deluxetable}{ccrrr}
\footnotesize
\tablecaption{Normal Points for Program Clusters}
\tablewidth{0pt}
\tablehead{
\colhead{$V$} & \colhead{\vmi} & \colhead{$n_1$} &
\colhead{$n_2$} & \colhead{$n_3$} }
\startdata
\multicolumn{5}{c}{a. Palomar~3} \nl
 20.884$\,\pm\,$0.034 & 0.978$\,\pm\,$0.024 &   3 &   3 &   3 \nl
 21.526$\,\pm\,$0.045 & 0.908$\,\pm\,$0.014 &   3 &   3 &   3 \nl
 21.742$\,\pm\,$0.019 & 0.919$\,\pm\,$0.004 &   7 &   7 &   7 \nl
 21.936$\,\pm\,$0.026 & 0.887$\,\pm\,$0.012 &   3 &   3 &   3 \nl
 22.119$\,\pm\,$0.016 & 0.890$\,\pm\,$0.005 &  10 &   9 &   8 \nl
 22.295$\,\pm\,$0.030 & 0.885$\,\pm\,$0.007 &   6 &   6 &   6 \nl
 22.525$\,\pm\,$0.023 & 0.847$\,\pm\,$0.019 &   7 &   7 &   7 \nl
 22.690$\,\pm\,$0.023 & 0.862$\,\pm\,$0.008 &   4 &   4 &   4 \nl
 22.927$\,\pm\,$0.022 & 0.807$\,\pm\,$0.014 &  10 &   9 &   9 \nl
 23.099$\,\pm\,$0.011 & 0.777$\,\pm\,$0.008 &  36 &  34 &  33 \nl
 23.299$\,\pm\,$0.010 & 0.676$\,\pm\,$0.005 &  34 &  31 &  30 \nl
 23.508$\,\pm\,$0.010 & 0.629$\,\pm\,$0.003 &  47 &  44 &  40 \nl
 23.690$\,\pm\,$0.010 & 0.622$\,\pm\,$0.002 &  48 &  44 &  43 \nl
 23.919$\,\pm\,$0.007 & 0.630$\,\pm\,$0.003 &  70 &  67 &  64 \nl
 24.103$\,\pm\,$0.006 & 0.627$\,\pm\,$0.003 &  87 &  85 &  81 \nl
 24.295$\,\pm\,$0.007 & 0.639$\,\pm\,$0.002 &  99 &  93 &  89 \nl
 24.501$\,\pm\,$0.005 & 0.652$\,\pm\,$0.002 & 122 & 115 & 109 \nl
 24.705$\,\pm\,$0.006 & 0.674$\,\pm\,$0.003 & 117 & 111 & 107 \nl
 24.908$\,\pm\,$0.006 & 0.701$\,\pm\,$0.004 & 117 & 114 & 111 \nl
 25.093$\,\pm\,$0.005 & 0.720$\,\pm\,$0.004 & 142 & 138 & 132 \nl
 25.304$\,\pm\,$0.005 & 0.738$\,\pm\,$0.004 & 153 & 147 & 139 \nl
 25.501$\,\pm\,$0.005 & 0.764$\,\pm\,$0.003 & 180 & 170 & 162 \nl
 25.698$\,\pm\,$0.005 & 0.800$\,\pm\,$0.004 & 204 & 193 & 183 \nl
 25.894$\,\pm\,$0.004 & 0.830$\,\pm\,$0.004 & 180 & 172 & 164 \nl
 26.100$\,\pm\,$0.004 & 0.867$\,\pm\,$0.005 & 184 & 177 & 171 \nl
 26.299$\,\pm\,$0.005 & 0.913$\,\pm\,$0.006 & 167 & 157 & 147 \nl
 26.502$\,\pm\,$0.005 & 0.955$\,\pm\,$0.006 & 163 & 155 & 149 \nl
 26.686$\,\pm\,$0.005 & 0.997$\,\pm\,$0.007 & 131 & 126 & 119 \nl
 26.893$\,\pm\,$0.005 & 1.048$\,\pm\,$0.009 &  96 &  91 &  85 \nl
 27.091$\,\pm\,$0.011 & 1.037$\,\pm\,$0.018 &  34 &  33 &  33 \nl
 27.257$\,\pm\,$0.016 & 1.093$\,\pm\,$0.014 &  17 &  16 &  15 \nl
 27.493$\,\pm\,$0.062 & 1.157$\,\pm\,$0.013 &   3 &   3 &   3 \nl
\multicolumn{5}{c}{b. Palomar~4} \nl
 20.119$\,\pm\,$0.064 & 1.082$\,\pm\,$0.008 &   3 &   3 &   3 \nl
 20.504$\,\pm\,$0.025 & 1.025$\,\pm\,$0.007 &   6 &   6 &   6 \nl
 20.663$\,\pm\,$0.022 & 1.007$\,\pm\,$0.010 &   5 &   5 &   5 \nl
 21.127$\,\pm\,$0.023 & 0.967$\,\pm\,$0.005 &   4 &   4 &   4 \nl
 21.305$\,\pm\,$0.043 & 0.968$\,\pm\,$0.018 &   4 &   4 &   4 \nl
 21.492$\,\pm\,$0.031 & 0.950$\,\pm\,$0.005 &   4 &   4 &   4 \nl
 22.095$\,\pm\,$0.019 & 0.907$\,\pm\,$0.004 &   7 &   7 &   7 \nl
 22.327$\,\pm\,$0.029 & 0.892$\,\pm\,$0.002 &   6 &   6 &   6 \nl
 22.441$\,\pm\,$0.021 & 0.893$\,\pm\,$0.000 &   3 &   3 &   3 \nl
 22.680$\,\pm\,$0.020 & 0.877$\,\pm\,$0.005 &  12 &  11 &  11 \nl
 22.914$\,\pm\,$0.020 & 0.885$\,\pm\,$0.009 &  12 &  11 &  10 \nl
 23.090$\,\pm\,$0.015 & 0.837$\,\pm\,$0.008 &  16 &  15 &  14 \nl
 23.331$\,\pm\,$0.011 & 0.817$\,\pm\,$0.006 &  22 &  20 &  19 \nl
 23.508$\,\pm\,$0.006 & 0.724$\,\pm\,$0.006 &  67 &  66 &  66 \nl
 23.714$\,\pm\,$0.007 & 0.644$\,\pm\,$0.002 &  61 &  58 &  55 \nl
 23.901$\,\pm\,$0.007 & 0.617$\,\pm\,$0.003 &  70 &  67 &  66 \nl
 24.098$\,\pm\,$0.006 & 0.621$\,\pm\,$0.002 &  96 &  93 &  89 \nl
 24.299$\,\pm\,$0.006 & 0.620$\,\pm\,$0.003 & 109 & 103 &  99 \nl
 24.499$\,\pm\,$0.005 & 0.632$\,\pm\,$0.003 & 132 & 129 & 124 \nl
 24.699$\,\pm\,$0.005 & 0.645$\,\pm\,$0.002 & 159 & 151 & 143 \nl
 24.908$\,\pm\,$0.004 & 0.665$\,\pm\,$0.003 & 167 & 160 & 153 \nl
 25.095$\,\pm\,$0.005 & 0.682$\,\pm\,$0.003 & 165 & 161 & 151 \nl
 25.300$\,\pm\,$0.004 & 0.706$\,\pm\,$0.002 & 224 & 209 & 195 \nl
 25.500$\,\pm\,$0.004 & 0.726$\,\pm\,$0.003 & 229 & 215 & 206 \nl
 25.698$\,\pm\,$0.004 & 0.753$\,\pm\,$0.003 & 227 & 212 & 201 \nl
 25.908$\,\pm\,$0.004 & 0.787$\,\pm\,$0.004 & 205 & 190 & 182 \nl
 26.099$\,\pm\,$0.004 & 0.826$\,\pm\,$0.005 & 195 & 184 & 172 \nl
 26.288$\,\pm\,$0.004 & 0.845$\,\pm\,$0.006 & 176 & 166 & 159 \nl
 26.502$\,\pm\,$0.004 & 0.893$\,\pm\,$0.008 & 159 & 151 & 144 \nl
 26.698$\,\pm\,$0.005 & 0.956$\,\pm\,$0.009 & 137 & 132 & 126 \nl
\multicolumn{5}{c}{c. Eridanus} \nl
 20.887$\,\pm\,$0.026 & 1.003$\,\pm\,$0.023 &   4 &   4 &   4 \nl
 21.339$\,\pm\,$0.022 & 0.935$\,\pm\,$0.013 &   3 &   3 &   3 \nl
 21.467$\,\pm\,$0.021 & 0.923$\,\pm\,$0.014 &   6 &   6 &   6 \nl
 21.701$\,\pm\,$0.024 & 0.913$\,\pm\,$0.006 &   8 &   7 &   7 \nl
 21.882$\,\pm\,$0.039 & 0.876$\,\pm\,$0.016 &   6 &   6 &   6 \nl
 22.130$\,\pm\,$0.023 & 0.899$\,\pm\,$0.012 &   4 &   4 &   4 \nl
 22.284$\,\pm\,$0.034 & 0.878$\,\pm\,$0.010 &   5 &   5 &   5 \nl
 22.524$\,\pm\,$0.018 & 0.857$\,\pm\,$0.006 &   9 &   8 &   8 \nl
 22.711$\,\pm\,$0.018 & 0.842$\,\pm\,$0.008 &   7 &   6 &   6 \nl
 22.886$\,\pm\,$0.022 & 0.818$\,\pm\,$0.008 &  12 &  11 &  11 \nl
 23.102$\,\pm\,$0.010 & 0.698$\,\pm\,$0.007 &  35 &  34 &  34 \nl
 23.303$\,\pm\,$0.009 & 0.638$\,\pm\,$0.004 &  40 &  40 &  40 \nl
 23.485$\,\pm\,$0.007 & 0.624$\,\pm\,$0.004 &  54 &  53 &  52 \nl
 23.710$\,\pm\,$0.007 & 0.615$\,\pm\,$0.004 &  59 &  57 &  54 \nl
 23.919$\,\pm\,$0.007 & 0.621$\,\pm\,$0.004 &  66 &  64 &  60 \nl
 24.094$\,\pm\,$0.007 & 0.636$\,\pm\,$0.004 &  82 &  79 &  76 \nl
 24.299$\,\pm\,$0.006 & 0.646$\,\pm\,$0.004 &  83 &  80 &  77 \nl
 24.496$\,\pm\,$0.006 & 0.664$\,\pm\,$0.003 & 125 & 119 & 111 \nl
 24.690$\,\pm\,$0.006 & 0.682$\,\pm\,$0.004 & 104 &  97 &  93 \nl
 24.888$\,\pm\,$0.006 & 0.711$\,\pm\,$0.003 & 113 & 107 & 101 \nl
 25.092$\,\pm\,$0.006 & 0.731$\,\pm\,$0.005 & 112 & 105 &  98 \nl
 25.307$\,\pm\,$0.005 & 0.747$\,\pm\,$0.006 & 113 & 110 & 107 \nl
 25.502$\,\pm\,$0.005 & 0.790$\,\pm\,$0.005 & 130 & 123 & 117 \nl
 25.701$\,\pm\,$0.005 & 0.814$\,\pm\,$0.006 & 113 & 109 & 104 \nl
 25.899$\,\pm\,$0.005 & 0.841$\,\pm\,$0.007 & 141 & 134 & 129 \nl
 26.108$\,\pm\,$0.005 & 0.906$\,\pm\,$0.007 & 115 & 109 & 104 \nl
 26.300$\,\pm\,$0.006 & 0.942$\,\pm\,$0.009 & 127 & 121 & 118 \nl
 26.502$\,\pm\,$0.006 & 0.998$\,\pm\,$0.009 & 115 & 111 & 107 \nl
 26.706$\,\pm\,$0.007 & 1.035$\,\pm\,$0.011 &  91 &  86 &  84 \nl
 26.891$\,\pm\,$0.006 & 1.048$\,\pm\,$0.010 &  85 &  81 &  80 \nl
 27.092$\,\pm\,$0.007 & 1.043$\,\pm\,$0.014 &  65 &  62 &  59 \nl
 27.274$\,\pm\,$0.014 & 1.093$\,\pm\,$0.025 &  20 &  19 &  18 \nl
 27.472$\,\pm\,$0.027 & 1.109$\,\pm\,$0.021 &   7 &   7 &   7 \nl
\multicolumn{5}{c}{d. \m3} \nl
 15.031$\,\pm\,$0.018 & 0.993$\,\pm\,$0.004 &   3 &   3 &   3 \nl
 16.343$\,\pm\,$0.014 & 0.882$\,\pm\,$0.005 &   6 &   6 &   6 \nl
 16.433$\,\pm\,$0.017 & 0.881$\,\pm\,$0.005 &   3 &   3 &   3 \nl
 16.660$\,\pm\,$0.004 & 0.874$\,\pm\,$0.017 &   4 &   4 &   4 \nl
 16.931$\,\pm\,$0.020 & 0.846$\,\pm\,$0.008 &   4 &   4 &   4 \nl
 17.736$\,\pm\,$0.014 & 0.810$\,\pm\,$0.007 &   3 &   3 &   3 \nl
 17.931$\,\pm\,$0.010 & 0.812$\,\pm\,$0.007 &   5 &   5 &   5 \nl
 18.034$\,\pm\,$0.009 & 0.799$\,\pm\,$0.007 &   8 &   8 &   8 \nl
 18.235$\,\pm\,$0.007 & 0.780$\,\pm\,$0.006 &   8 &   7 &   7 \nl
 18.363$\,\pm\,$0.007 & 0.743$\,\pm\,$0.008 &  19 &  18 &  18 \nl
 18.455$\,\pm\,$0.006 & 0.680$\,\pm\,$0.009 &  20 &  20 &  20 \nl
 18.552$\,\pm\,$0.007 & 0.647$\,\pm\,$0.005 &  23 &  22 &  21 \nl
 18.655$\,\pm\,$0.009 & 0.608$\,\pm\,$0.011 &  14 &  14 &  14 \nl
 18.744$\,\pm\,$0.005 & 0.610$\,\pm\,$0.003 &  34 &  32 &  30 \nl
 18.865$\,\pm\,$0.005 & 0.583$\,\pm\,$0.003 &  30 &  29 &  29 \nl
 18.953$\,\pm\,$0.006 & 0.585$\,\pm\,$0.002 &  41 &  39 &  37 \nl
 19.048$\,\pm\,$0.004 & 0.581$\,\pm\,$0.002 &  44 &  41 &  40 \nl
 19.151$\,\pm\,$0.005 & 0.584$\,\pm\,$0.002 &  53 &  50 &  49 \nl
 19.250$\,\pm\,$0.003 & 0.581$\,\pm\,$0.002 &  59 &  54 &  51 \nl
 19.346$\,\pm\,$0.004 & 0.581$\,\pm\,$0.002 &  53 &  50 &  45 \nl
 19.447$\,\pm\,$0.004 & 0.589$\,\pm\,$0.002 &  53 &  49 &  49 \nl
 19.545$\,\pm\,$0.004 & 0.590$\,\pm\,$0.002 &  60 &  57 &  54 \nl
 19.656$\,\pm\,$0.004 & 0.606$\,\pm\,$0.003 &  56 &  52 &  50 \nl
 19.752$\,\pm\,$0.004 & 0.609$\,\pm\,$0.003 &  55 &  53 &  49 \nl
 19.848$\,\pm\,$0.003 & 0.614$\,\pm\,$0.002 &  66 &  61 &  56 \nl
 19.954$\,\pm\,$0.004 & 0.628$\,\pm\,$0.003 &  67 &  65 &  61 \nl
 20.057$\,\pm\,$0.004 & 0.638$\,\pm\,$0.002 &  73 &  69 &  65 \nl
 20.149$\,\pm\,$0.004 & 0.646$\,\pm\,$0.003 &  74 &  69 &  63 \nl
 20.247$\,\pm\,$0.003 & 0.659$\,\pm\,$0.002 &  80 &  75 &  72 \nl
 20.349$\,\pm\,$0.003 & 0.672$\,\pm\,$0.003 &  78 &  77 &  74 \nl
 20.454$\,\pm\,$0.003 & 0.687$\,\pm\,$0.002 &  82 &  75 &  70 \nl
 20.546$\,\pm\,$0.004 & 0.702$\,\pm\,$0.003 &  97 &  90 &  86 \nl
 20.646$\,\pm\,$0.003 & 0.715$\,\pm\,$0.003 &  93 &  91 &  84 \nl
 20.754$\,\pm\,$0.004 & 0.734$\,\pm\,$0.004 &  89 &  88 &  82 \nl
 20.850$\,\pm\,$0.004 & 0.749$\,\pm\,$0.003 &  64 &  59 &  55 \nl
 20.950$\,\pm\,$0.003 & 0.760$\,\pm\,$0.003 &  95 &  90 &  84 \nl
 21.053$\,\pm\,$0.004 & 0.777$\,\pm\,$0.003 &  85 &  81 &  77 \nl
 21.149$\,\pm\,$0.003 & 0.796$\,\pm\,$0.003 &  75 &  72 &  67 \nl
 21.247$\,\pm\,$0.003 & 0.813$\,\pm\,$0.005 &  67 &  65 &  61 \nl
 21.352$\,\pm\,$0.004 & 0.829$\,\pm\,$0.003 &  58 &  55 &  53 \nl
 21.444$\,\pm\,$0.004 & 0.847$\,\pm\,$0.004 &  61 &  56 &  52 \nl
 21.552$\,\pm\,$0.005 & 0.875$\,\pm\,$0.003 &  49 &  46 &  44 \nl
 21.649$\,\pm\,$0.004 & 0.895$\,\pm\,$0.004 &  49 &  45 &  42 \nl
 21.753$\,\pm\,$0.005 & 0.914$\,\pm\,$0.004 &  53 &  50 &  47 \nl
 21.849$\,\pm\,$0.004 & 0.941$\,\pm\,$0.005 &  49 &  47 &  43 \nl
 21.937$\,\pm\,$0.005 & 0.966$\,\pm\,$0.006 &  44 &  41 &  39 \nl
 22.058$\,\pm\,$0.004 & 0.987$\,\pm\,$0.006 &  49 &  47 &  43 \nl
 22.142$\,\pm\,$0.005 & 1.009$\,\pm\,$0.005 &  41 &  38 &  36 \nl
 22.254$\,\pm\,$0.005 & 1.033$\,\pm\,$0.006 &  35 &  32 &  31 \nl
 22.343$\,\pm\,$0.005 & 1.055$\,\pm\,$0.005 &  39 &  36 &  34 \nl
 22.450$\,\pm\,$0.004 & 1.078$\,\pm\,$0.004 &  46 &  42 &  41 \nl
 22.550$\,\pm\,$0.005 & 1.113$\,\pm\,$0.008 &  31 &  29 &  27 \nl
 22.637$\,\pm\,$0.008 & 1.183$\,\pm\,$0.015 &  17 &  16 &  16 \nl
 22.758$\,\pm\,$0.009 & 1.166$\,\pm\,$0.021 &  10 &  10 &  10 \nl
 22.830$\,\pm\,$0.009 & 1.178$\,\pm\,$0.019 &   6 &   6 &   6 \nl
\multicolumn{5}{c}{e. \m5} \nl
 14.954$\,\pm\,$0.021 & 1.004$\,\pm\,$0.012 &   3 &   3 &   3 \nl
 15.039$\,\pm\,$0.036 & 0.996$\,\pm\,$0.003 &   3 &   3 &   3 \nl
 15.570$\,\pm\,$0.016 & 0.948$\,\pm\,$0.013 &   4 &   4 &   4 \nl
 15.645$\,\pm\,$0.032 & 0.945$\,\pm\,$0.009 &   3 &   3 &   3 \nl
 15.854$\,\pm\,$0.021 & 0.930$\,\pm\,$0.011 &   4 &   4 &   4 \nl
 15.965$\,\pm\,$0.016 & 0.928$\,\pm\,$0.011 &   4 &   4 &   4 \nl
 16.432$\,\pm\,$0.013 & 0.894$\,\pm\,$0.012 &   5 &   5 &   5 \nl
 16.557$\,\pm\,$0.017 & 0.890$\,\pm\,$0.006 &   5 &   5 &   5 \nl
 16.678$\,\pm\,$0.007 & 0.888$\,\pm\,$0.005 &   4 &   4 &   4 \nl
 16.758$\,\pm\,$0.005 & 0.873$\,\pm\,$0.005 &   5 &   5 &   5 \nl
 16.948$\,\pm\,$0.009 & 0.877$\,\pm\,$0.008 &   8 &   8 &   8 \nl
 17.042$\,\pm\,$0.013 & 0.869$\,\pm\,$0.010 &   7 &   7 &   7 \nl
 17.156$\,\pm\,$0.005 & 0.865$\,\pm\,$0.004 &  12 &  12 &  12 \nl
 17.283$\,\pm\,$0.005 & 0.834$\,\pm\,$0.009 &   3 &   3 &   3 \nl
 17.361$\,\pm\,$0.015 & 0.848$\,\pm\,$0.011 &   6 &   6 &   6 \nl
 17.432$\,\pm\,$0.011 & 0.858$\,\pm\,$0.008 &   5 &   5 &   5 \nl
 17.558$\,\pm\,$0.010 & 0.828$\,\pm\,$0.007 &  10 &  10 &  10 \nl
 17.653$\,\pm\,$0.009 & 0.830$\,\pm\,$0.007 &  11 &  11 &  11 \nl
 17.751$\,\pm\,$0.011 & 0.809$\,\pm\,$0.013 &   8 &   7 &   7 \nl
 17.856$\,\pm\,$0.007 & 0.767$\,\pm\,$0.006 &  26 &  26 &  26 \nl
 17.952$\,\pm\,$0.006 & 0.717$\,\pm\,$0.005 &  34 &  31 &  30 \nl
 18.050$\,\pm\,$0.006 & 0.674$\,\pm\,$0.003 &  27 &  25 &  23 \nl
 18.148$\,\pm\,$0.007 & 0.641$\,\pm\,$0.002 &  30 &  27 &  26 \nl
 18.248$\,\pm\,$0.006 & 0.630$\,\pm\,$0.003 &  36 &  32 &  31 \nl
 18.361$\,\pm\,$0.004 & 0.624$\,\pm\,$0.003 &  38 &  37 &  36 \nl
 18.453$\,\pm\,$0.005 & 0.631$\,\pm\,$0.003 &  46 &  45 &  42 \nl
 18.537$\,\pm\,$0.006 & 0.624$\,\pm\,$0.005 &  26 &  25 &  24 \nl
 18.645$\,\pm\,$0.004 & 0.627$\,\pm\,$0.002 &  58 &  55 &  53 \nl
 18.749$\,\pm\,$0.004 & 0.631$\,\pm\,$0.003 &  49 &  46 &  45 \nl
 18.847$\,\pm\,$0.004 & 0.633$\,\pm\,$0.003 &  50 &  49 &  46 \nl
 18.958$\,\pm\,$0.005 & 0.641$\,\pm\,$0.003 &  48 &  46 &  44 \nl
 19.055$\,\pm\,$0.003 & 0.648$\,\pm\,$0.003 &  50 &  49 &  48 \nl
 19.158$\,\pm\,$0.005 & 0.650$\,\pm\,$0.002 &  49 &  48 &  45 \nl
 19.250$\,\pm\,$0.004 & 0.661$\,\pm\,$0.003 &  56 &  55 &  53 \nl
 19.352$\,\pm\,$0.003 & 0.668$\,\pm\,$0.003 &  74 &  70 &  69 \nl
 19.450$\,\pm\,$0.004 & 0.675$\,\pm\,$0.002 &  60 &  58 &  54 \nl
 19.551$\,\pm\,$0.004 & 0.688$\,\pm\,$0.002 &  89 &  86 &  82 \nl
 19.658$\,\pm\,$0.004 & 0.698$\,\pm\,$0.003 &  50 &  48 &  47 \nl
 19.751$\,\pm\,$0.004 & 0.709$\,\pm\,$0.003 &  69 &  67 &  64 \nl
 19.851$\,\pm\,$0.004 & 0.719$\,\pm\,$0.003 &  65 &  62 &  61 \nl
 19.955$\,\pm\,$0.004 & 0.742$\,\pm\,$0.003 &  60 &  56 &  54 \nl
 20.049$\,\pm\,$0.004 & 0.746$\,\pm\,$0.003 &  85 &  81 &  77 \nl
 20.154$\,\pm\,$0.004 & 0.765$\,\pm\,$0.003 &  65 &  63 &  60 \nl
 20.255$\,\pm\,$0.004 & 0.780$\,\pm\,$0.003 &  63 &  62 &  60 \nl
 20.345$\,\pm\,$0.004 & 0.783$\,\pm\,$0.004 &  58 &  56 &  52 \nl
 20.448$\,\pm\,$0.003 & 0.809$\,\pm\,$0.003 &  66 &  63 &  62 \nl
 20.545$\,\pm\,$0.004 & 0.829$\,\pm\,$0.004 &  62 &  59 &  57 \nl
 20.649$\,\pm\,$0.004 & 0.840$\,\pm\,$0.003 &  64 &  60 &  58 \nl
 20.744$\,\pm\,$0.004 & 0.857$\,\pm\,$0.005 &  60 &  57 &  54 \nl
 20.852$\,\pm\,$0.002 & 0.875$\,\pm\,$0.003 &  68 &  65 &  63 \nl
 20.946$\,\pm\,$0.004 & 0.898$\,\pm\,$0.005 &  46 &  45 &  44 \nl
 21.044$\,\pm\,$0.005 & 0.916$\,\pm\,$0.004 &  54 &  51 &  48 \nl
 21.140$\,\pm\,$0.004 & 0.930$\,\pm\,$0.004 &  47 &  44 &  42 \nl
 21.258$\,\pm\,$0.005 & 0.969$\,\pm\,$0.005 &  42 &  40 &  40 \nl
 21.356$\,\pm\,$0.005 & 0.985$\,\pm\,$0.005 &  37 &  35 &  32 \nl
 21.441$\,\pm\,$0.004 & 1.002$\,\pm\,$0.004 &  44 &  42 &  40 \nl
 21.552$\,\pm\,$0.005 & 1.036$\,\pm\,$0.006 &  36 &  34 &  34 \nl
 21.660$\,\pm\,$0.005 & 1.058$\,\pm\,$0.006 &  32 &  30 &  29 \nl
 21.746$\,\pm\,$0.006 & 1.073$\,\pm\,$0.010 &  19 &  18 &  17 \nl
 21.858$\,\pm\,$0.007 & 1.095$\,\pm\,$0.009 &  23 &  22 &  21 \nl
 21.958$\,\pm\,$0.006 & 1.162$\,\pm\,$0.007 &  27 &  25 &  24 \nl
 22.058$\,\pm\,$0.007 & 1.168$\,\pm\,$0.010 &  17 &  17 &  17 \nl
 22.134$\,\pm\,$0.007 & 1.186$\,\pm\,$0.013 &  19 &  18 &  17 \nl
 22.251$\,\pm\,$0.011 & 1.275$\,\pm\,$0.017 &  11 &  11 &  11 \nl
 22.346$\,\pm\,$0.015 & 1.244$\,\pm\,$0.043 &   3 &   3 &   3 \nl
 22.537$\,\pm\,$0.023 & 1.350$\,\pm\,$0.024 &   3 &   3 &   3 \nl
\enddata
\end{deluxetable}

\clearpage
\begin{deluxetable}{lcccccc}
\tablecaption{Turnoff Locations for Program Clusters}
\tablecolumns{7}
\tablewidth{0pt}
\tablehead{
\colhead{Cluster} & \colhead{$V_{\small MSTO}$} & \colhead{$(\vmi)_{\small MSTO}$} & 
\colhead{$V^-_{+0.10}$} & \colhead{$V^-_{+0.05}$} & \colhead{$V^+_{+0.05}$}
& \colhead{$V^+_{+0.10}$}
}
\startdata
Palomar~3 & 23.83 & 0.614 & 23.22 & 23.35 & 24.63 & 25.09 \nl
Palomar~4 & 24.12 & 0.614 & 23.52 & 23.64 & 24.93 & 25.38 \nl
Eridanus  & 23.70 & 0.610 & 23.09 & 23.21 & 24.50 & 24.95 \nl
\m3       & 19.16 & 0.578 & 18.53 & 18.67 & 19.97 & 20.40 \nl
\m5       & 18.56 & 0.614 & 17.95 & 18.09 & 19.37 & 19.84 \nl
\enddata
\end{deluxetable}
 
\clearpage
\begin{deluxetable}{lrrccc}
\tablecaption{Giant-Branch Offsets for Program Clusters}
\tablecolumns{6}
\tablewidth{0pt}
\tablehead{
\colhead{} & \colhead{mean} & \colhead{median} & \colhead{standard}
&& \colhead{median} \nl
\colhead{Cluster} & \colhead{offset} & \colhead{offset} & 
\colhead{deviation} & \colhead{N} & \colhead{$\sigma(\vmi)$} 
}
\startdata
Palomar~3 & --$\,$0.0035 & -0.001 & 0.021 & 45 & 0.013 \nl
Palomar~4 & --$\,$0.0037 & --$\,$0.003 & 0.012 & 51 & 0.011 \nl
Eridanus  & +0.0078 & +0.009 & 0.020 & 46 & 0.014 \nl
\enddata
\end{deluxetable}

\clearpage
\begin{deluxetable}{lrrccc}
\tablecaption{Giant-Branch Offsets for Program Clusters}
\tablecolumns{6}
\tablewidth{0pt}
\tablehead{
\colhead{} & \colhead{mean} & \colhead{median} & \colhead{standard}
&& \colhead{median} \nl
\colhead{Cluster} & \colhead{offset} & \colhead{offset} & 
\colhead{deviation} & \colhead{N} & \colhead{$\sigma(\vmi)$} 
}
\startdata
Palomar~3 & +0.0096 & +0.015 & 0.020 & 44 & 0.013 \nl
Palomar~4 & +0.0123 & +0.012 & 0.010 & 50 & 0.011 \nl
Eridanus  & +0.0205 & +0.026 & 0.022 & 45 & 0.014 \nl
\m3       & --$\,$0.0139 & --$\,$0.015 & 0.014 & 46 & 0.005 \nl
\m5       & --$\,$0.0073 & --$\,$0.006 & 0.019 & 85 & 0.010 \nl
\enddata
\end{deluxetable}

\clearpage
\begin{deluxetable}{lcccccc}
\tablecaption{Fiducial Parameters for Program Clusters}
\tablecolumns{7}
\tablewidth{0pt}
\tablehead{
\colhead{Cluster} & \colhead{[Fe/H]} & \colhead{$E(\bmv)$} & 
\colhead{$(\vmi)_{\circ,\small TO}$} &
\colhead{$V_{\small HB}$} & \colhead{$V_{\small MSTO}$} & \colhead{$\Delta V$}
}
\startdata
Palomar~3 & $-1.57$ & 0.03 & 0.579 & 20.51 & 23.83 & 3.32 \nl
Palomar~4 & $-1.28$ & 0.01 & 0.595 & 20.80 & 24.12 & 3.32 \nl
Eridanus  & $-1.42$ & 0.02 & 0.590 & 20.42 & 23.70 & 3.28 \nl
\m3       & $-1.57$ & 0.01 & 0.578 & 15.68 & 19.14 & 3.46 \nl
\m5       & $-1.29$ & 0.03 & 0.579 & 15.15 & 18.59 & 3.44 \nl
\enddata
\end{deluxetable}

\end{document}